\documentclass[nofootinbib, twocolumn, longbibliography,superscriptaddress]{revtex4-1}
\usepackage{amsmath}
\usepackage{amsthm}
\usepackage{amssymb}
\usepackage{graphicx}
\usepackage{hyperref}
\usepackage{color}
\usepackage{array}
\usepackage[makeroom]{cancel}
\usepackage[capitalize]{cleveref}
\usepackage{changes}
\usepackage{braket}
\usepackage{bbm}
\usepackage{bm}
\usepackage{booktabs}
\usepackage{titlesec}
\usepackage{comment}
\usepackage{subfigure}
\usepackage{adjustbox}

\newcommand{\beq}{\begin{equation}}
\newcommand{\eeq}{\end{equation}}

\usepackage{multirow}
\usepackage{cellspace} 
\usepackage{algorithm}
\usepackage{algpseudocode}

\definecolor{MyDarkBlue}{rgb}{0,0.1,0.75}
\hypersetup{pdfborder={0 0 0},colorlinks,breaklinks=true,
  urlcolor={MyDarkBlue},citecolor={MyDarkBlue},linkcolor={MyDarkBlue}
}

\def\ket#1{\left| #1 \right\rangle}

\theoremstyle{plain}

\theoremstyle{definition}

\theoremstyle{remark}

\usepackage{nameref}
\newcommand{\figref}[1]{Fig.~\ref{#1}}

\newcommand{\tr}{{\rm Tr}}

\newcommand{\ketbra}[2]{\vert #1 \rangle \langle #2 \vert}

\algrenewcommand\algorithmicrequire{\textbf{Input:}}
\algrenewcommand\algorithmicensure{\textbf{Output:}}

\newcommand{\bea}{\begin{eqnarray}}
\newcommand{\eea}{\end{eqnarray}}
\newcommand{\be}{\begin{equation}}
\newcommand{\ee}{\end{equation}}
\newcommand{\ba}{\begin{equation}\begin{aligned}}
\newcommand{\ea}{\end{aligned}\end{equation}}
\newcommand{\eqs}[1]{\begin{align}#1\end{align}}

\def\1{\mathds{1}}

\def\mB{\mathcal{B}}

\def\mE{\mathcal{E}}
\def\mF{\mathcal{F}}

\def\mM{\mathcal{M}}
\def\mN{\mathcal{N}}
\def\mO{\mathcal{O}}

\def\mU{\mathcal{U}}

\def\({\left(}
\def\){\right)}
\def\[{\left[}
\def\]{\right]}

\renewcommand{\theparagraph}{\arabic{paragraph}.}

\titleformat{\paragraph}[runin]
  {\normalfont\itshape}
  {\theparagraph}
  {0.4em}
  {}
  
 \titlespacing{\paragraph}
  {10pt}{0.2\baselineskip}{0.3\baselineskip}

\newcommand{\review}[1]{{\textcolor{black}{#1}}}

\begin{document}
\title{Physics-Inspired Extrapolation for efficient error mitigation and hardware certification}

\author{Pablo Díez-Valle} 
\email{pvalle@itg.es}
\affiliation{LG Electronics Toronto AI Lab, Toronto, Ontario M5V 1M3, Canada} 
\affiliation{Instituto de Física Fundamental IFF-CSIC, Calle Serrano 113b, Madrid 28006, Spain} 
\affiliation{Instituto Tecnológico de Galicia, Cantón Grande 9, Planta 3, 15003 A Coruña, Spain}

\author{Gaurav Saxena} 
\email[(Corresponding author) ]{gaurav.saxena@lge.com}
\affiliation{LG Electronics Toronto AI Lab, Toronto, Ontario M5V 1M3, Canada}

\author{Jack S. Baker}
\affiliation{LG Electronics Toronto AI Lab, Toronto, Ontario M5V 1M3, Canada} 

\author{Jun-Ho Lee}
\affiliation{LG Electronics, AI Lab, CTO Div, 19, Yangjae-daero 11-gil, Seocho-gu, Seoul, 06772, Republic of Korea} 

\author{Thi Ha Kyaw}
\email{thiha.kyaw@lge.com}
\affiliation{LG Electronics Toronto AI Lab, Toronto, Ontario M5V 1M3, Canada}

\begin{abstract}
Quantum error mitigation (QEM) is essential for the noisy intermediate-scale quantum era, and will remain relevant for early fault-tolerant quantum computers, where logical error rates are still significant.
However, most QEM methods incur an exponential sampling overhead to achieve unbiased estimates, limiting their practical applicability.
Recently, error mitigation by restricted evolution (EMRE) was shown to estimate expectation values with constant sampling overhead, albeit with a small bias that grows with circuit size and noise level.
Building upon the EMRE framework, here, we propose physics-inspired extrapolation (PIE), a linear circuit runtime protocol that achieves enhanced accuracy without incurring substantial overhead. 
Unlike traditional heuristic zero-noise extrapolation methods, PIE provides a theoretical foundation for the operational interpretation of its fitting parameters and demonstrates convergence to unbiased estimates as noise decreases. 
Distinctively, the slope of the extrapolation fit corresponds to the max-relative entropy between the ideal and noisy circuits, enabling quantitative hardware certification alongside error mitigation, with no additional computational overhead.
We also demonstrate the efficacy of this method on IBMQ hardware and apply it to simulate 84-qubit quantum dynamics efficiently. 
Our results show that PIE yields accurate, low-variance error mitigated estimates, establishing it as a practical and scalable strategy for both error mitigation and hardware certification for near-term and early fault-tolerant quantum computers.
\end{abstract}
\maketitle

\section{Introduction}

\noindent The ubiquitous presence of noise in various quantum hardware modalities presents a fundamental obstacle in realizing an ultimate quantum machine capable of executing industry-relevant applications.
Examples of such applications include simulating FeMoco molecules \cite{Lee2021Jul}, modeling spin defects \cite{Baker2024Sep}, designing solid-state battery materials \cite{Delgado2022Sep}, optimizing solar cell design \cite{Motlagh2024Nov}, and accelerating drug discovery \cite{Santagati2024Apr}, tasks that remain intractable for even the current best classical computers~\cite{Hoefler2023Apr,Lee2023Apr}.

To overcome the above challenge and achieve fault tolerance in quantum computers, quantum error correction (QEC) \cite{BibEntry2013Sep} emerges as the ultimate tool. 
QEC refers to a set of techniques that protect quantum information against errors arising due to both environmental effects (such as decoherence) and imperfect gate operations (such as random bit-flip error). 
Despite the promise of fault-tolerant quantum computing, today’s quantum processors remain constrained to the noisy intermediate-scale quantum (NISQ) regime~\cite{Preskill2018Aug, Bharti2022Feb}, characterized by a limited number of qubits and susceptibility to various noise sources. Implementing fully error-corrected quantum algorithms \cite{Gidney2025May} capable of outperforming classical computations remains infeasible due to prohibitive resource overheads, with each logical qubit requiring hundreds to thousands of physical qubits. 
Consequently, until fully fault-tolerant quantum systems are realized, we must shift our focus to alternative strategies \cite{Bharti2022Feb} for enhancing computational accuracy on existing hardware.

\begin{figure*}[th]
        \centering
        \includegraphics[width=1\linewidth]{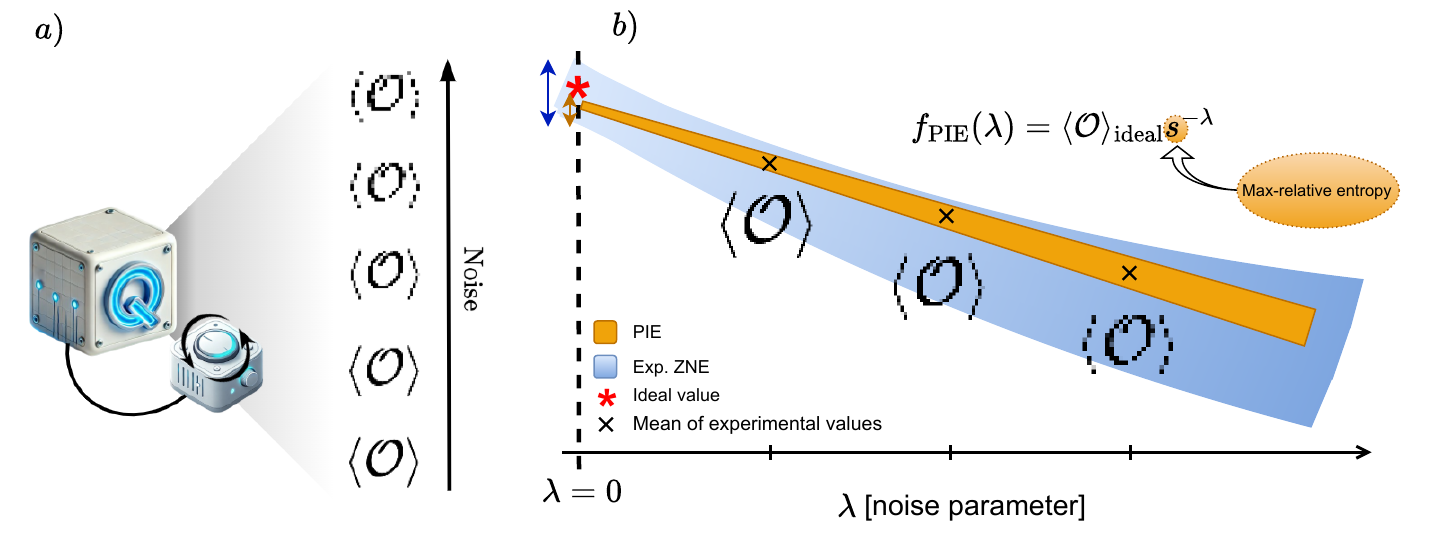}
        \caption{Workflow and Performance Comparison of the PIE Method.
        (a) Expectation values $\langle \mathcal{O} \rangle$ are measured from a quantum circuit executed on a noisy quantum processor. Measurements are repeated at different effective noise levels, achieved by inserting identity-equivalent operations (e.g., gate folding) into the circuit. 
        (b) Extrapolation to the zero-noise limit using the collected data. The yellow curve shows the \textit{PIE} fit, defined as 
        $f_{\text{PIE}}(\lambda) = \langle \mathcal{O} \rangle_{\text{ideal}} s^{-\lambda}$, 
        where $s$ is the max relative entropy between the noisy and ideal operations. This is compared to \textit{Exponential Zero-Noise Extrapolation (Exp. ZNE)} (blue region). PIE exhibits three key advantages:
        (i) Exponentially reduced data requirements to achieve accurate extrapolation, 
        (ii) Lower variance in the inferred zero-noise observable, as evidenced by the narrower confidence region, 
        (iii) Physically meaningful fit parameters, including the parameter $s$, interpreted as the maximum relative entropy, which can then be used to certify quantum circuit performance (see the main text.)
    }
        \label{fig:emre_key_flowchart}
\end{figure*}
One particularly promising approach is quantum error mitigation (QEM)~\cite{RevModPhys.95.045005, Li2017Jun,Temme2017Nov,Endo2018Jul,Bonet-Monroig2018Dec,Giurgica-Tiron2020Oct, Sun2021Mar, Shaw2021May, Lowe2021Jul, Koczor2021Sep, Wang2021MitigatingQE, Suzuki2022Mar, Bultrini2023Jun, liu2024virtualchannelpurification, SK2024emre}. Unlike QEC, QEM techniques do not require the costly overhead of encoding logical qubits. 
They aim to suppress or eliminate the impact of noise through post-processing of measurement outcomes, enabling more accurate estimation of quantum observables even within the limitations of current NISQ devices.
These methods are algorithmic schemes designed to reduce the bias (absolute error between the obtained and ideal results) in the expectation value, and as such do not have a universal or a unique definition. 
Among the most widely used QEM methods is zero-noise extrapolation (ZNE)~\cite{PhysRevX.7.021050,Temme2017Nov}, where noise is treated as a parameter.
By systematically amplifying the noise in the quantum circuit, one can obtain a series of noisy expectation values and extrapolate them back to the zero-noise limit, thereby estimating the ideal result.
Another prominent method is the probabilistic error cancellation (PEC)~\cite{Temme2017Nov}, which provides unbiased estimates of the expectation value by using quasi-probabilistic decomposition of the gates in the circuit. However, the unbiased estimate is achieved using a sampling overhead that scales exponentially with the number of gates in the circuit, rendering it impractical for large circuits.

Although current noisy quantum hardware lack enough high-quality physical qubits to construct a sufficient number of logical qubits for algorithmic relevance, an intermediate “early fault-tolerant (EFT)” era is anticipated before full fault-tolerance is achieved. In this EFT era, processors may offer a few hundred to a few thousand logical qubits enabled by imperfect error correction schemes.
Noise will not be fully eliminated, and algorithms will still suffer performance losses. Consequently, error mitigation will remain crucial, complementing partial or limited error correction to mitigate against logical noise \cite{PRXQuantum.3.010345, dutkiewicz2025errormitigationcircuitdivision}. 
This approach is expected to produce the highest possible performance using EFT devices. 
Thus, error mitigation schemes that do not incur large computational overhead will play a pivotal role in maximizing efficiency and accuracy of the outcomes during the EFT era.

Here, we introduce an error mitigation technique based on the constant runtime Error Mitigation by Restricted Evolution (EMRE) protocol that we recently proposed~\cite{SK2024emre}.
The primary contribution of this manuscript is the development and experimental demonstration of the EMRE-inspired extrapolation, termed Physics-Inspired Extrapolation (PIE), on IBMQ devices, with a benchmarking against standard resource-efficient ZNE approaches.
While the protocol is mathematically equivalent to a specific choice of a ZNE method, we show that the PIE technique exhibits four principal advantages (see~\figref{fig:emre_key_flowchart}).
First, parameters in the extrapolation function used in PIE have a clear operational interpretation, which is often missing in conventional extrapolation-based mitigation schemes; second, the computational runtime scales linearly with respect to the number of data points chosen for extrapolation, making it computationally efficient; third, the associated estimation variance remains low, ensuring reliability in the results; and fourth, and most distinctively, PIE enables 
direct quantification of the distance between the ideal and noisy circuits in terms of maximum relative entropy, evaluating noise impact and enabling hardware certification.
We demonstrate that in the low-noise regime, PIE method provides accurate estimates of the expectation values using only linear runtime, making it a strong candidate for deployment on EFT quantum processors, where hardware resources and error rates remain significant constraints.
Additionally, we also develop another error mitigation protocol, the inverse EMRE, which leverages hardware noise characterization to obtain the generalized quasi-probabilistic decomposition (GQPD) of each gate in the circuit to perform EMRE.
Owing to noise characterization and the computation of optimal GQPD, the inverse EMRE method has comparatively longer runtime and larger deviations than the PIE method, and so, we have presented the full discussion of the inverse EMRE method, including its implementation and numerical and experimental results, in the supplementary information. 
\\

\section{Physics-Inspired Extrapolation (PIE)}
\label{sec_PIE}
Characterizing noise in physical hardware is a challenging and resource-intensive task. 
Since error mitigation techniques aim to estimate the ideal expectation values, procedures that avoid any extra computational cost, such as those incurred from noise characterization, are generally preferred. 
In this scheme of things, ZNE has proven to be very practical~\cite{Kim2023Jun}.
ZNE works by systematically increasing the circuit fault rate, getting the erroneous expectation value at the different fault rates, and then extrapolating the expectation values to the zero-noise limit.
To enhance circuit fault rate to implement ZNE, one of the widely used methods is through circuit folding, where identities are added to the circuit, increasing the length of the circuit and systematically increasing noise~\cite{Giurgica-Tiron2020Oct, majumdar2023zne}. 
For example, if we assume that the whole circuit is represented by a unitary operator $U$, then folding the circuit once would mean that we implement the circuit $U\circ U^{\dagger}\circ U$ which, in ideal non-erroneous circumstances, should yield the same result as the original case. 
However, due to the faulty nature of the gates, circuit folding increases the errors in the circuit.
By folding the circuit multiple times, we can get erroneous results as a function of the circuit fault rate or the number of circuit foldings.
The expectation values obtained in the procedure are then extrapolated to the zero-noise limit.

An uncertainty that ZNE presents is in the choice of the extrapolation function.
For example, in~\cite{PhysRevX.7.021050} a linear extrapolation function was employed to approximate the ideal expectation value, while~\cite{Temme2017Nov} demonstrated that the ideal expectation value can be approximated using a polynomial function for small circuit fault rates, whereas~\cite{PhysRevX.8.031027} showed that exponential extrapolation offers improved bias compared to other extrapolation techniques. Other extrapolation functions used for error mitigation can be found in the review article~\cite{RevModPhys.95.045005}.
In addition, for extrapolations going beyond Richardson extrapolation, there are no analytical expressions for bias or variance in most cases~\cite{RevModPhys.95.045005}.
Furthermore, no operational meaning has been given to the extrapolation methods. Intuitive explanations on why the exponential extrapolation method works have been provided, but no solid rigorous explanation/reasoning has been given regarding which exponential function should be used for extrapolation.

Here, we give a resolution to the above by introducing the error mitigation approach PIE, which is free from heuristic methods. 
Each step in the protocol, including the choice of extrapolation fit, is guided by operational meaning.
Using the extrapolation-based techniques and the EMRE framework, PIE can efficiently estimate the expectation value without requiring explicit knowledge of the underlying hardware noise.
We determine an extrapolation function that enables an accurate prediction of the ideal (noiseless) expectation value. Accuracy improves considerably in low-noise regimes, as theoretical approximations converge toward exact values as the probability of noise decreases (see Eq.~\eqref{eq_expectdvalueapprox}). 
We validate the effectiveness of PIE through extensive numerical simulations and experimental demonstrations on IBMQ hardware, showing that it yields reliable, noise-robust estimates of expectation values.
We benchmark our results against standard ZNE methods using various extrapolation models, including linear, quadratic, and exponential fits.
We find that the expectation value estimate using our PIE method yields highly accurate results while avoiding the large standard deviations often observed with exponential ZNE.
Thus, we demonstrate that PIE provides a practically motivated and efficient error mitigation technique, bringing the quantum utility era closer to reality.
Importantly, our results do not contradict the fundamental limitations established in \cite{Quek2024Oct}. 
At high noise probabilities, the estimates from PIE begin to deviate from the ideal values as PIE does not address the underlying source of exponential bias arising from the omitted terms in EMRE.
However, as gate fidelities continue to improve, PIE becomes increasingly effective, offering a practical and efficient error mitigation strategy that uses a non-heuristic and practically motivated extrapolation fit whose parameters have an underlying operational interpretation.
In the following, we discuss how PIE leverages the EMRE framework and circuit folding to get the extrapolation function and estimate the expectation value efficiently.

EMRE relies on the GQPD of an ideal gate and mitigates errors by restricting the decomposition to its positive component, thereby approximating the ideal operation~\cite{SK2024emre}.
To achieve minimum bias, it is necessary to employ the optimal GQPD.
However, obtaining this optimal decomposition requires hardware noise characterization, which in itself is both a time-consuming and a resource-intensive process.
Even if a (not necessarily optimal) GQPD is known, EMRE can be implemented using only a constant sample overhead.
This constant sample overhead comes at the cost of a finite bias which grows exponentially with the circuit size.
Since our objective is to mitigate errors without incurring significant runtime or bias, we need to find a way to circumvent noise characterization.
In order to do that, we propose constructing a GQPD for each ideal gate (and thus for the ideal circuit) using the noisy gates. 
We show below that it is always possible to construct a GQPD (not necessarily optimal) of the ideal unitary operation from erroneous ones using maximum relative entropy.

The max-relative entropy between a unitary operation $\mU$ and its noisy version $\mE\circ\mU$ where $\mE$ denotes the noise channel, is given by
\eqs{D_{\max}(\mU\|\mE\circ\mU)&= \log \min \{s: s\mE\circ\mU \geq \mU,\, s\geq 1\}\\
&= \log \min \{s:  sJ^{\mE\circ\mU} \geq J^{\mU},\, s\geq 1\}\label{eq:s_definition}}
where $J^{\mU}:= {\rm id}\otimes \mU(\Phi^+_{\tilde{A}A})$ represents the Choi matrix of $\mU$, $\Phi^+_{\tilde{A}A}$ is the unnormalized maximally entangled state on the system $\tilde{A}A$ where $|\tilde{A}|=|A|$, and the inequality in the above equations implies that the matrix $sJ^{\mE\circ\mU} - J^{\mU}$ is a positive semidefinite matrix.
Denoting $\log s_{\rm opt}$ to be the max-relative entropy between the two operations, we can express the ideal operation $\mU$ as
\eqs{\mU = s_{\rm opt}\mE\circ\mU - (s_{\rm opt} - 1)\mM \label{eq_udecompo}}
where $\mM$ is some quantum channel.
Note that if $\mU$ denotes an ideal unitary circuit, then $\mE$ can be considered as the effective noise channel defined as $\mE =\mF\circ \mU^{\dagger}$ where $\mF$ is a quantum channel representing the noisy circuit.
Interestingly, PIE offers a hidden advantage by not requiring the computation of the GQPD for any gate in the circuit and using their erroneous implementations instead. In standard EMRE, the GQPD of all the gates in the circuit must be known, and the original circuit is replaced with a modified one sampled according to the GQPD associated with each gate.
Here, only the gates affected by noise are effectively approximated by their noisy counterparts. As a result, the overall product of the generalized robustness of all gates remains small, as we see from our experimental and numerical results, which further contributes to a reduced bias.
However, the associated $s$ remains unknown, but we later show that it can be determined using PIE.

Given the above, it is clear that a GQPD can be constructed for an ideal unitary operation in terms of its erroneous version. 
This proof of existence of a GQPD is enough to perform error mitigation as we describe below, and we need not find the optimal coefficient $s_{\rm opt}$ or the channel $\mM$.

Since EMRE works by approximating the ideal unitary operation to the positive part of its GQPD, we can approximate the expectation value of an observable, say $\mO$, as follows \eqs{\langle\mO\rangle_{\mU}\approx s\langle\mO\rangle_{\mE\circ \mU} \label{eq_expectdvalueapprox}}
where $\langle\mO\rangle_{\mM}:= \tr[\mO \mM(\ketbra{0}{0})]$ for any quantum channel, $\mM$.
Note that the smaller the noise probability, the closer the RHS in the above equation will be to the ideal expectation value in the LHS.
As a remark, we also emphasize that from the above approximation relation, one might incorrectly infer that $s$ depends on the choice of observable. As we show below, for any nontrivial observable $\mO$, the slope of the extrapolation fit (which depends solely on $s$) remains invariant as long as the circuit and noise are fixed, implying that $s$ is independent of the observable.
Indeed, since $s$ is a distance measure, it must remain independent of the observable and depend only on the ideal and noisy operations, as defined in Eq.~\ref{eq:s_definition}.

Now, to enhance the errors in the circuit, we apply global circuit folding.
We assume that the same noise acts throughout the circuit even after increasing the circuit depth, and note that the max relative entropy between $\mU^{\dagger}$ and $\mE\circ\mU^{\dagger}$ is the same as the max relative entropy between $\mU$ and $\mE\circ\mU$.
So, after $n$ circuit foldings, we can approximate the expectation value of $\mO$ as follows:
\eqs{\label{expectation_after_folding}\langle\mO\rangle_{\mU} = \langle\mO\rangle_{\mU\circ(\mU^{\dagger}\circ\mU)^n}\approx s^{(2n+1)} \langle\mO\rangle_{\mE\circ \mU\circ (\mE\circ \mU^{\dagger}\circ \mE\circ \mU)^n}\,.}
Next, by rearranging the terms in Eq.~\eqref{expectation_after_folding}, we get the following general expression
\eqs{ \langle\mO\rangle_{\rm err}\approx \frac{\langle\mO\rangle_{\rm ideal}}{s^{2n+1}}
}
where $\langle\mO\rangle_{\rm err}$ represents the erroneous expectation value after $n$ circuit foldings and $\langle\mO\rangle_{\rm ideal}$ denotes the ideal expectation value.
Using the above formula, if we plot the erroneous expectation values against the number of circuit folding, we can extrapolate it back to the zero-noise case.
If we let $\lambda = 2n+1$, and denote the experimental expectation values as a function of $\lambda$, i.e., $f(\lambda) = \langle \mO \rangle_{\rm err}$, we get the following extrapolation function to be used:
\eqs{f(\lambda) = \langle\mO\rangle_{\rm ideal}s^{-\lambda}\,, \label{eq_PIEexponential}}
where $\lambda=0$ indicates the case without error. Further, taking the log on both sides, we can cast the above expression as
\eqs{\log\left(f\right) = \log\left(\langle\mO\rangle_{\rm ideal}\right)-\lambda\log(s)\,, \label{eq_PIEfunction}}
which is a linear extrapolation between the logarithm of the obtained experimental values on the $y$-axis against the number of circuit folding on the $x$-axis. 
Then, the intercept on the $y$-axis at $\lambda=0$ gives us the logarithm of the ideal expectation value estimate.
Eq.~\eqref{eq_PIEfunction} allows the use of linear regression to calculate $\langle\mO\rangle_{\rm ideal}$ and $s$, which is numerically stable and less sensitive to initialization issues than nonlinear least-squares methods.
If the expectation values are negative, then one can take the log of their absolute values, do the extrapolation, get the estimate and multiply the result by -1. We have demonstrated this in our quantum chemistry examples (see Fig.~\ref{fig_chemistryexperiment}). 

\begin{figure}[h]
\includegraphics[scale=0.52]{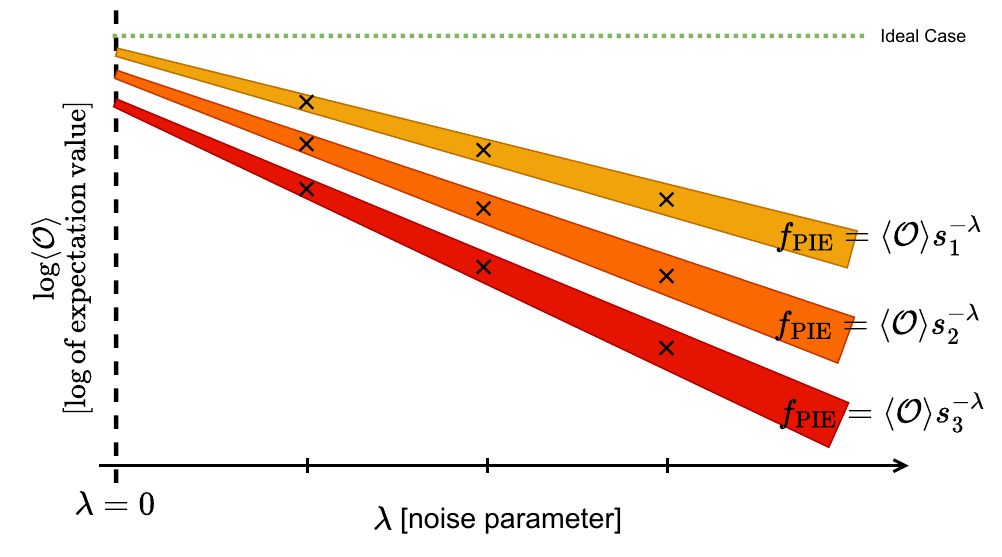}
\caption{\label{fig_MaxRelativeEntropy} (Color online) Illustration of estimating expectation values using PIE on different quantum hardware. The slope of each curve reflects the hardware's sensitivity to noise (yellow (least steep), orange, and red (steepest)), enabling resource-efficient certification via max-relative entropy without prior knowledge of the noise or ideal circuit.}
\end{figure}

We also emphasize that PIE can, in principle, address SPAM errors, though doing so introduces an additional approximation error if no complementary mitigation methods are employed. SPAM errors can be separated into two components: incorrect state preparation and incorrect measurement.
We can view state preparation as part of the quantum circuit itself. In other words, if we only require that the qubits are initialized in the $|0\rangle^{\otimes n}$ state perfectly, then state preparation may be regarded as a pre-computation step whose imperfections simply contribute to the overall gate-level noise.
For measurement errors, one way is to use alternate/complimentary EM method such as T-Rex which is widely used in practice and has worked well.
Alternatively, if no such complimentary EM method is used, one may decompose the ideal observable (similar to GQPD) into a linear combination of implementable observable/s (which will form the positive component of the decomposition) and an additional matrix (the negative part), in line with what we did for other operations in the circuit (see for reference, Eq.~\eqref{eq_udecompo}).
Next, we can approximate the ideal observable $\mO$ to be the erroneous observable $\mO^{\rm err}$ scaled by a factor, say $s_m$, i.e., approximate $\mO$ as $\mO\approx s_m\mO^{\rm err}$.
Since folding cannot be applied to this observable, $s_m$ cannot be absorbed into the max-relative entropy of the circuit $s$.
Consequently, we will have an extra parameter in Eqs.~\eqref{eq_expectdvalueapprox} and~\eqref{expectation_after_folding}, and the ideal outcome after $n$ circuit folds can be approximated as $\langle \mO\rangle_{\rm ideal}\approx s_m s^{(2n+1)}\langle \mO\rangle_{\rm err}\,.$
Taking logarithms introduces an additional parameter, which must be learned.
This parameter can be estimated using simple observables within a learning-based procedure and subsequently applied to the noisy measurements of the target observable to recover its ideal expectation value.

A notable feature of the extrapolation in Eq.\eqref{eq_PIEfunction} is that the slope of the linear curve corresponds to the logarithm of the maximum relative entropy between the ideal and the noisy circuits, thereby quantifying their operational distance.
The max-relative entropy serves as a measure of distinguishability and is widely used in quantum information theory, particularly in tasks such as quantum channel discrimination~\cite{PhysRevResearch.1.033169,li2022sequentialquantumchanneldiscrimination,  Kossmann2024Oct,  Fang2025Jul} and resource quantification~\cite{Datta_IEEE_min_max,  PhysRevLett.119.150405,PhysRevLett.123.150401, Saxena2020Jun, PhysRevLett.124.090505, PhysRevA.106.042422, RevModPhys.91.025001}.

In our setting, the max-relative entropy serves as a measure of how far the noisy circuit deviates from its ideal counterpart. 
In the broader literature, estimating the distance between ideal and noisy channels is known to be a challenging task. 
Tomography techniques, such as quantum process tomography, allow one to fully characterize the quantum channel but become impractically expensive as system size grows, due to its exponential scaling~\cite{Poyatos_complete_1997,mohseni_quantum_2008,torlai_quantum_2023}. Fidelity estimation protocols, such as randomized benchmarking, offer scalable alternatives by estimating average gate performance through statistical overlap between ideal and noisy states~\cite{knill_randomized_2008,dankert_exact_2009,helsen_general_2022}. 
Yet, all these methods typically require some assumptions or knowledge of the noise or ideal channel's action on the input states in order to compare and reconstruct the process~\cite{eisert_quantum_2020}.
In contrast, PIE offers a resource-efficient approach to estimate this distance via the max relative entropy without requiring prior knowledge of the ideal circuit. 
This perspective naturally frames PIE not only as an error-mitigation protocol but also as a quantum circuit certification tool that operates in parallel with expectation-value estimation, without incurring additional overhead.
Such dual-purpose functionality aligns with emerging trends in quantum system certification, where protocols aim to validate circuit integrity alongside computational output~\cite{PRXQuantum.2.010201}.

As an illustration,~\figref{fig_MaxRelativeEntropy} presents three curves corresponding to data collected from three distinct quantum hardware. Each curve plots the extrapolated expectation value of the same observable, with the ideal value indicated by the green dotted line.
The slope of each curve reflects the hardware's sensitivity to noise: shallower slopes indicate greater resilience. Specifically, the yellow curve exhibits the least negative slope, suggesting minimal degradation under noise and thus superior hardware performance. The orange and red curves show progressively steeper slopes, indicating increased susceptibility to noise.
This behavior enables a form of hardware certification, where the slope serves as a proxy for the maximum relative entropy between the ideal and noisy operations. In other words, a smaller slope implies a closer match to the ideal circuit. Notably, this certification is achieved without additional experimental overhead, as it runs in parallel with expectation value estimation. 

\begin{figure*}[th]
\includegraphics[width=0.8\linewidth]{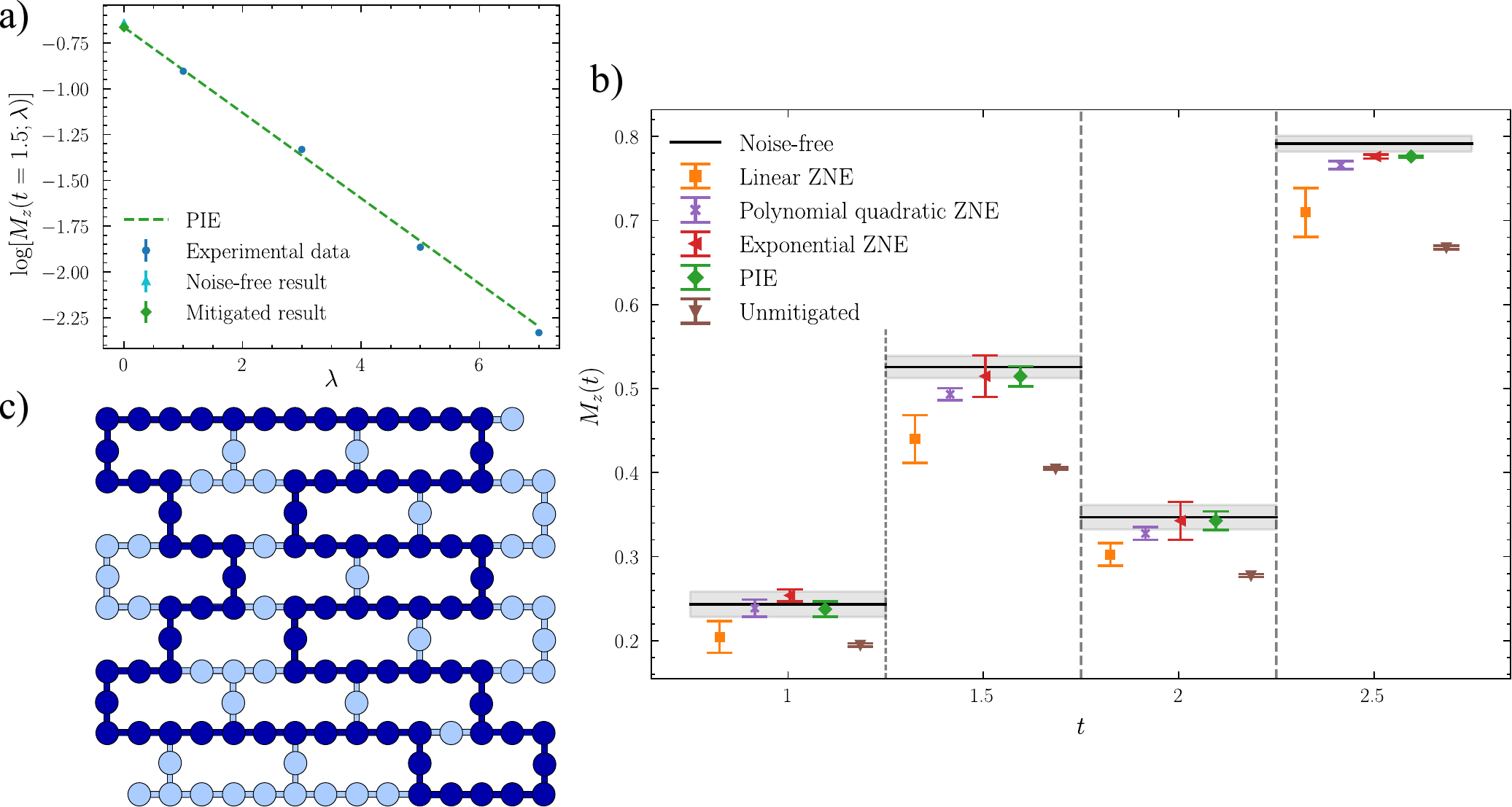}
\caption{\label{fig_experimentalresults} (Color online) Experimental error-mitigated global magnetization $M_z$ on the two Trotter steps simulation of the one-dimensional Ising chain with $N=84$ spins. (a) PIE extrapolation from the noise amplified expectation values for $t=1.5$. \review{The noise-free, experimental unmitigated, and PIE mitigated results are $0.526\pm0.013$, $0.405\pm0,002$, and $0.515\pm0.012$, respectively.} (b) Comparison of extrapolation methods. (c) Topology of the IBM Eagle superconducting processor ibm kyiv with the ring of qubits selected for the experiments in dark blue. All experiments were performed using four extrapolation points and 4096 shots.}
\end{figure*}

\section{Results}
The PIE method enables accurate and efficient error mitigation without requiring explicit noise characterization.
In this section, we focus on benchmarking PIE against the standard ZNE techniques and report the corresponding results.
As a proof-of-concept, we primarily use the transverse-field Ising model to benchmark our results. 
The main reason behind using this model is that it is one of the simplest yet non-trivial many-body models exhibiting quantum phase transition~\cite{Sachdev_2011, Dutta_Aeppli_Chakrabarti_Divakaran_Rosenbaum_Sen_2015}, is exactly solvable in one-dimension~\cite{PFEUTY197079, heyl2014dynamical}, and captures the physics of some real magnetic materials~\cite{doi:10.1126/science.1180085, PhysRevResearch.1.033141,PhysRevLett.123.067203,Wang2018Feb, PhysRevB.96.024439, Faure2018Jul, OKUTANI2015779,Kyaw2020Jan}.
We also benchmark PIE using quantum chemistry simulations, in particular, by calculating the ground-state energies of the hydrogen (H$_2$) and lithium hydride (LiH) molecules.
These results (and corresponding experimental setup) are presented in the subsections below.
The application of PIE in much more complex chemical molecular systems will be shown in our subsequent study \cite{Delmar2025}.

\subsection{Ising model with transverse-field}
To benchmark the PIE method against various ZNE methods, let us look at the dynamical simulation of the ubiquitous one-dimensional quantum Ising spin chain Hamiltonian with periodic boundary condition ($\hbar=1$ throughout),
\begin{equation}
    {H} = -\sum_{i=0}^{N-2} \sigma^x_{i}\sigma^x_{i+1} - \sigma^x_{N-1}\sigma^x_{0} -  \Gamma \sum_{i=0}^{N-1} \sigma^z_{i} \,,
    \label{eq_Isingchain}
\end{equation}
with $N$ representing the total number of spins or qubits, $\Gamma=0.5$ being the transverse field strength\review{, and $\sigma^x$, $\sigma^z$ denoting the usual Pauli matrices}.
The time-evolution is approximated via the first-order Suzuki-Trotter decomposition,
\begin{equation}
\ket{\Psi(t)} = e^{-i {H} t} \ket{\Psi(0)} \approx U_{R} \ket{\Psi(0)}\,,
\label{eq_trotter}
\end{equation}
with
\begin{equation}
    U_{R} = \prod_{j=0}^R \left[e^{i(t/R)\sum_{i=1}^{N-2} \sigma^x_{i}\sigma^x_{i+1}}e^{i(t/R) \sigma^x_{N-1}\sigma^x_{0} }e^{i\Gamma(t/R)\sum_{i=1}^N \sigma^z_{i}}\right]\,,
    \label{eq_trotterevol}
\end{equation}
where $\ket{\Psi(0)}$ is an initial state of the quantum system, $j$ denotes each Trotter step, and the time-evolution operator is being decomposed into a total of $R$ Trotter steps.
In the limit $R\rightarrow\infty$, the simulation becomes exact.

We are interested in estimating the global transverse field magnetization $M_z(t)$, defined as
\begin{equation}
    M_z(t) = \sum_{i=0}^{N-1}\dfrac{\tr[\sigma^z_i U_R\ketbra{\Psi(0)}{\Psi(0)}U_R^{\dagger}]}{N}\,,
    \label{eq_globalmagnetization}
\end{equation} 
given the noisy hardware and various QEM techniques at hand.
We compute its evolution at different times $t$, starting from an initial state in which all spins are aligned in the positive $\sigma_z$ direction such that $M_z(0)=1$.
While the one-dimensional Ising model described in Eq.~\eqref{eq_Isingchain} admits analytical solutions for an arbitrary number of spins \cite{Suzuki,Auerbach}, our primary focus is not on the simulation error introduced by the Trotter approximation in Eq.~\eqref{eq_trotter}. Instead, we focus on evaluating the performance of error mitigation algorithms when applied to actual hardware demonstrations and theoretical studies under realistic and commonly encountered noise models, such as depolarizing, dephasing, and random Pauli noises.
In this respect, the noise-free output of the Trotter circuit in Eq.~\eqref{eq_trotter} is taken as the ideal result regardless of the number of Trotter steps.  
We benchmark the PIE method against different extrapolation-based error mitigation techniques by deploying them to estimate the global transverse magnetization for different evolution times.
The noise amplification is carried out by \textit{circuit folding}~\cite{Giurgica-Tiron2020Oct}, where the trotterized time-evolution operator Eq.~\eqref{eq_trotterevol} is replaced by
\begin{equation}
    U_R \rightarrow U_R\left(U_R^{\dagger}U_R\right)^n\,, 
\end{equation}
where $n$ is a non-negative integer that corresponds to the number of unitary circuit foldings. 
This transformation does not modify the noise-free expectation values of any observable, while it allows us to amplify the noise in the quantum circuit. 
Noise amplification is parameterized by the scaling factor $\lambda = 1 + 2n$, which can intuitively be interpreted as the effective noise level corresponding to the number of folded circuit layers.
The $\lambda = 0$ case formally corresponds to the ideal, noise-free circuit.
The error mitigation approach consists of computing the expectation value of the observable at different noise levels
\begin{equation}
    M_z(t;\lambda) = f(\lambda;c_0,\ldots,c_m)\,,
\end{equation}
where $(c_0,\ldots,c_m)$ are free model parameters that must be adjusted to fit the experimental data.
Then, by using a function $f$, one is to extrapolate the ideal noise-free value at $\lambda=0$, resulting in a mitigated estimate of the expectation value (see \figref{fig:emre_key_flowchart}(b)).
The variance associated with this estimate is determined from the variance of the fitting parameters.
The resulting expectation value and its variance are given as
\begin{equation}
    M_z(t;0) = f(0)  \,\,\,,\,\,\,\textnormal{Var}[M_z(t;0)]=\bm J \bm \Sigma \bm J^T\,,
\end{equation}
where $\bm J$ and $\bm \Sigma$ denote the Jacobian and the covariance matrix, respectively. As explained in section~\ref{sec_PIE}, our proposed PIE technique leads to the following linear extrapolation model (Eq.\eqref{eq_PIEfunction}):
\eqs{
    \log f_{\textnormal{PIE}}(\lambda) = c_0 + c_1 \lambda\,.
\label{eq_PIEfunction}
}
From this, we obtain
\eqs{
    M_z(t;0) = e^{c_0} \,\,,\,\,\textnormal{Var}[M_z(t;0)] = e^{2c_0}\textnormal{Var}[c_0]\,,
}
where both the fitting parameters $c_0$ and $c_1$ carry physical meanings.
Specifically, $c_0$ corresponds to the logarithm of the noise-free expectation value, while $c_1$ represents the logarithm of the distance between the ideal and noisy operations, as quantified by the max-relative entropy.
The operational interpretation of the fitting parameters is typically absent in other extrapolation-based mitigation protocols, providing strong motivation for employing PIE over alternative methods relying on heuristic approximations.

\begin{figure*}[th]
\includegraphics[width=0.75\linewidth]{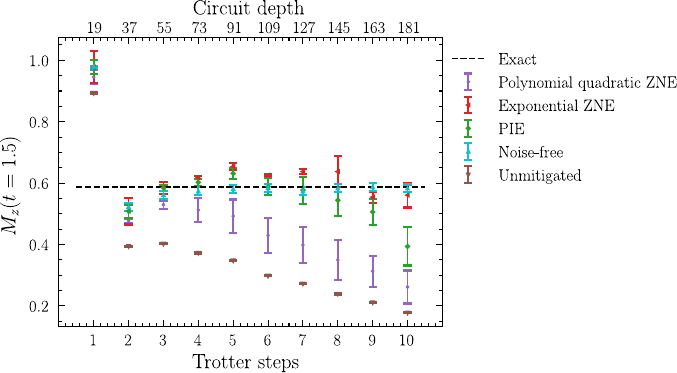}
\caption{\label{fig_incresaingtrottersteps} (Color online) PIE performance as we increase the number of Trotter steps and hence the depth of the quantum circuit and the noise in the computation of the global magnetization $M_z$. We report results from experiments on $N=84$ qubits of the IBM Eagle superconducting processor ibm kyiv (the selected qubits are displayed in Figure~\ref{fig_experimentalresults}). We also report the mitigated results obtained by other common extrapolation techniques. We observe that PIE is the most accurate protocol for low noise levels, while the results deviate from the noise-free and the exponential ZNE result for large circuit depths.}
\end{figure*}
We benchmark the performance of PIE with various standard ZNE methods, including linear, where the extrapolation function is given by
\eqs{f_{\textnormal{lin}}(\lambda) = l_0 + l_1 \lambda }
with the noise-free expectation value and its variance computed as
\eqs{M_z(t;0) = l_0 \:,\:\textnormal{Var}[M_z(t;0)] = \textnormal{Var}[l_0]\,.}
Similarly, for the quadratic polynomial extrapolation, the function used is of the form
\eqs{f_{\textnormal{quad}}(\lambda) = q_0 + q_1 \lambda + q_2 \lambda^2}
leading to
\eqs{M_z(t;0) = q_0 \:,\:{\rm Var}[M_z(t;0)] = {\rm Var}[q_0]\,,}
and exponential extrapolation, where the extrapolation function is of the following form,
\eqs{f_{\textnormal{exp}}(\lambda) = x_0 + x_1 e^{-x_2\lambda} }
resulting in
\eqs{M_z(t;0) &= x_0 + x_1 \:,\nonumber \\ {\rm Var}[M_z(t;0)] = \textnormal{Var}[x_0] &+ \textnormal{Var}[x_1] + 2\textnormal{Covar}[x_0,x_1]\,.}
In~\figref{fig_experimentalresults}, we report the actual device implementation of the error-mitigated global magnetization at different times $t = 1,\, 1.5,\, 2,$ and $2.5$, for an Ising chain \eqref{eq_trotterevol} with $N=84$ spins. 
The time dynamics simulation is approximated by a Trotterized evolution with $R=2$ Trotter steps, which require quantum circuits of depth 37 after transpilation. 
The experiments were carried out on the IBM Eagle superconducting processor, \textit{ibm kyiv}, using the set of connected qubits shown in~\figref{fig_experimentalresults}c.
Erroneous expectation values at different noise levels ($\lambda=1,3,5,7$) are computed from 4096 shots, with readout error mitigation applied via IBM's \textit{twirled readout error extinction} protocol. 
The noise-free benchmark is calculated from 4096 shots of the simulation of $20$-qubit circuits, since the global magnetization of the Hamiltonian in~\eqref{eq_Isingchain} is largely insensitive to system size, i.e., independent of the number of spins, beyond finite-size effects. 
Although the exact value of the observable from infinitely many shots can be computed at this system size, we use this benchmark because, even in the absence of errors, the ideal quantum circuit would provide a result from a finite number of shots.
The PIE, linear, and quadratic extrapolations allow for a least-squares polynomial fit, while the exponential extrapolation required a nonlinear least-squares fit.
We observe that the PIE estimates are consistent with the results of the standard ZNE methods, with exponential ZNE and PIE providing nominally better mitigated estimates. 
However, the variance of the mitigated expectation value associated with the nonlinear exponential ZNE fit is larger than that obtained by the PIE-based linear fit, highlighting the practical advantage of the PIE method in terms of statistical stability. 
We report similar results on the IBM Heron superconducting processor \textit{ibm torino} for an Ising chain with $N = 44$ spins (see the supplementary information).

We further assess the performance of our PIE method for increasing circuit depth, as presented in~\figref{fig_incresaingtrottersteps}. 
As before, we compute the error-mitigated global magnetization of an Ising chain with $N=84$ spins, using Suzuki-Trotter decompositions executed on the same set of qubits of the IBM Eagle superconducting processor \textit{ibm kyiv}. 
In this case, we kept the time constant at $t=1.5$ and varied the number of Trotter steps from 1 to 10. 
The mitigated results are obtained via extrapolation from expectation values computed at noise levels $\lambda=1,\,3,\,5,\, {\rm and}\, 7$ using 4096 shots and IBM’s twirled readout error extinction for readout mitigation, consistent with the previous experiment. 
The results demonstrate that PIE is the most accurate protocol for low to moderate circuit depths. However, at larger depths, the results deviate from the noise-free and exponential ZNE results.
In this regime, we concur that with growing circuit size, the contribution of the quantum channel $\mM$ in \eqref{eq_udecompo} becomes prominent, and the approximation in \eqref{eq_expectdvalueapprox} worsens substantially.
Despite this, it still continues to outperform the quadratic extrapolation in terms of accuracy.

\begin{figure*}[th]
\includegraphics[width=1.0\linewidth]{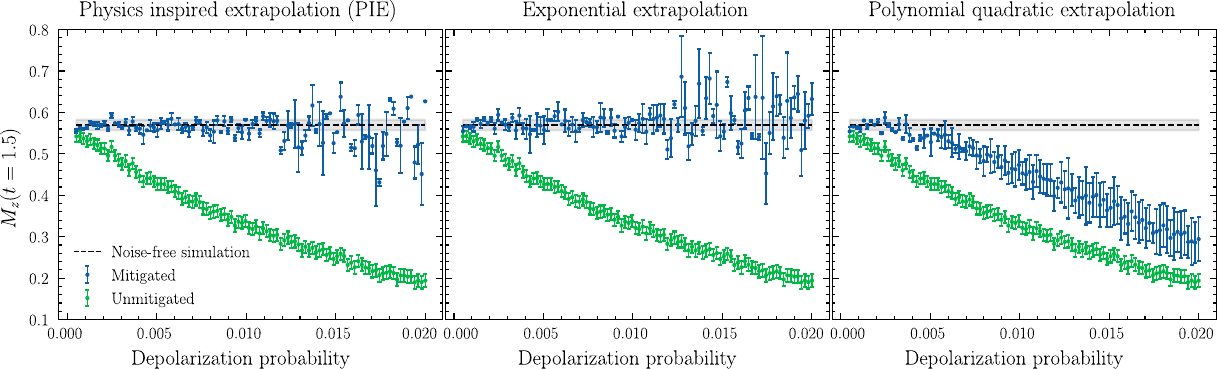}
\caption{\label{fig_depolnoise} (Color online) Numerical results of the global magnetization $M_z$ on the three Trotter steps simulation of the
one-dimensional Ising chain at time $t=1.5$ with N = 8 spins using a depolarizing noise model. We display the error mitigated results (dots) from three extrapolation techniques (PIE, exponential, and polynomial quadratic), together with the standard deviation of the results (bars). We also show the unmitigated or noisy result and the noise-free simulation, with the corresponding standard deviation (bars and shaded area, respectively). All experiments were performed using four extrapolation points and 4096 shots.}
\end{figure*}
In addition to the IBM Q implementation results, we benchmark the PIE method through numerical simulations by considering realistic noise models such as the depolarizing noise, dephasing noise, inhomogeneous Pauli noise and Pauli Lindblad noise. 
The simulation results under the depolarizing noise model are presented in \figref{fig_depolnoise}.
There, we analyze the behavior of PIE, exponential ZNE, and quadratic polynomial ZNE under the depolarizing quantum error channel given by
\begin{equation}
    \mE(\rho) = (1-\omega)\rho + \omega \tr[\rho]\frac{I}{2^n}\,, 
    \label{eq_depolnoise}
\end{equation}
where $\omega$ denotes the depolarization probability, $n$ is the number of qubits on which the channel acts, and $\rho$ denotes a quantum state on which the noise $\mE$ acts.
Specifically, we implement a model in which circuits are transpiled to Hadamard, Pauli-X, Pauli-Y, single-qubit Z-axis rotation, and the two-qubit controlled-NOT gates, all subject to the noise described in \eqref{eq_depolnoise}.
The numerical results of global magnetization $M_z$ with three Trotter steps (circuit depth $=27$) in the simulation of the one-dimensional Ising chain at time $t=1.5$ with N = 8 spins confirm the advantage of PIE in low-noise regimes. 
Similar and consistent PIE performance is observed for various other noise models, which can be seen in the supplementary material.\\

\subsection{Quantum Chemistry example}
To further validate PIE, we perform an additional experiment: calculating the ground-state energies of the H$_2$ and LiH molecules.
The second quantization form of the electronic Hamiltonian is given by
\begin{equation}
    H_{el} = \sum_{pq}h_{pq}a_{p}^{\dagger}a_{q} + 
    \frac{1}{2} \sum_{pqrs}h_{pqrs}a^{\dagger}_{p}a^{\dagger}_{q}a_{r}a_{s},
    \label{eq:2nd}
\end{equation}
where $a^{\dagger}_i(a_i)$ is the creation (annihilation) operator acting on $i$-th state, $h_{pq}$ are one-particle integrals given by
\begin{equation}
    h_{pq} = \int d\sigma \phi^*_p(\sigma)\left(-\frac{\nabla^2_r}{2} - \sum_i \frac{Z_i}{|R_i - r|}\right)\phi_q(\sigma)\,,
    \label{eq:1e}
\end{equation}
$h_{pqrs}$ are two-particle integrals given by
\begin{equation}
    h_{pqrs} = \int d\sigma_1 d\sigma_2 \frac{\phi^*_p(\sigma_1)\phi^*_q(\sigma_2)\phi_s(\sigma_1)\phi_r(\sigma_2)}{\left|r_1 - r_2 \right|} \,,
    \label{eq:1e2}
\end{equation}
$\phi_i$ is a basis used to represent the wave function, and $\sigma_i$ is a spatial and spin coordinate with $\sigma_i= (r_i; s_i)$~\cite{Yudong2019}. The first and second terms in parentheses in Eq.~\eqref{eq:1e} represent the kinetic and potential energy of electrons, respectively.

\begin{figure*}[th]
\includegraphics[width=1.0\linewidth]{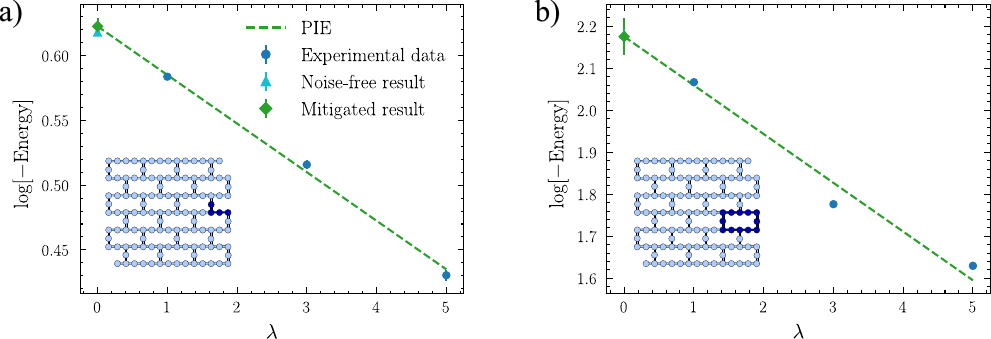}
\caption{\label{fig_chemistryexperiment} (Color online) Experimental error-mitigated results on the ground-state energy calculation of two distinct molecules. The experiments were performed on the IBM Eagle superconducting processor ibm sherbrooke, using 4096 shots of a hardware-efficient SU(2) 2-local ansatz with two-layer repetitions and optimal angles. Together with the mitigated result, we show the PIE extrapolation from the noise amplified expectation values, the result from a noise-free execution of the quantum circuit, and the topology of the quantum computer with the qubits selected for the experiments in dark blue. a) Hydrogen molecule H$_2$ with an equilibrium distance of 0.74 \text{\AA} (four qubit ansatz). b) Lithium hydride LiH with an equilibrium distance of 1.6 \text{\AA} (twelve qubit ansatz).}
\end{figure*}
We chose the STO-3G basis set and the equilibrium distance between two hydrogen atoms of 0.74 \text{\AA} and distance between Li and H as 1.60 \text{\AA}, and performed Hartree-Fock calculations to get $h_{pq}$ and $h_{pqrs}$ using \texttt{PySCF}~\cite{pyscf2018,pyscf2020}. We applied the Jordan-Wigner transformation to map the second-quantized electronic Hamiltonian in~\eqref{eq:2nd} to a qubit Hamiltonian ($\hat H_q$), which can be written in the form below
\eqs{\label{eq:qubit_hamiltonian}
\hat H_q = \sum_i \alpha_i \hat P_i
}
where $\hat P_i$ is a Pauli string and $\alpha_i$ its coefficient. H$_2$ and LiH molecules require 4 and 12 qubits, respectively.

We used the variational quantum eigensolver on a local simulator (a classical computer) to obtain the optimal parameters \{${\theta}_{opt}$\} of a parameterized quantum circuit, $|\Psi(\theta)\rangle=|U(\theta) \rangle$, that minimize the energy of the molecules ($E=\langle \Psi (\theta)|\hat H_q|\Psi(\theta) \rangle$). We employed a hardware-efficient SU(2) 2-local ansatz with two layers (each layer consisting of Pauli Y and Pauli Z parameterized single-qubit rotations and two-qubit CNOT entangling gates between nearest neighbors), and performed parameter update using the classical optimizer COBYLA~\cite{COBYLA}. 
We then calculate the molecule energy using 4096 shots of the quantum circuit with optimal parameters, $|\Psi\rangle\equiv|\Psi(\theta_{opt})\rangle$, executed on the IBM Eagle superconducting processor, \textit{ibm sherbrooke}.
The 4- and 12-qubit quantum circuits for the H$_2$ and LiH molecules lead to a depth of 25 and 53, respectively, after transpilation. 
Note that we are not interested in the capability of the hardware-efficient ansatz to approximate the ground state of the molecule, but in testing how PIE is able to compute the expected value of a noise-free execution of the ansatz with the most optimal parameters.
Moreover, unlike the global transverse-field magnetization in Eq.~\eqref{eq_globalmagnetization}, the energy of a fully mixed state for molecular Hamiltonians is generally non‑zero. This arises from the nuclear–nuclear repulsion term as well as the non‑vanishing trace of the electronic Hamiltonian. Consequently, since Eq.~\eqref{eq_PIEfunction} asymptotically approaches zero, the PIE fit should be regarded as an approximation in the short $\lambda$ regime.
In general, to work with a non-traceless Hamiltonian, one must shift the qubit Hamiltonian (as in Eq.~\eqref{eq:qubit_hamiltonian}) to the traceless part, obtain results using PIE and then shift it back to obtain the estimate of the ideal result.

For the quantum circuit and the Hamiltonian that we prepared, we calculated the expectation value, $E=\langle \Psi |\hat H_q|\Psi \rangle$, and its variance, $\text{Var} = \langle \Psi | \hat H_q^2 | \Psi \rangle - \langle \Psi | \hat H_q | \Psi \rangle^2$, at different noise levels ($\lambda=1,3,5$) to perform PIE. We found that PIE provides good error-mitigated results as shown in Fig.~\ref{fig_chemistryexperiment}. In particular, in the case of the 4 qubit H$_{2}$ molecule, the reference value electronic ground state energy is -1.852 Ha, the noise-free execution of the hardware-efficient ansatz with optimal parameters result is $-1.855\pm0.005$ Ha, the experimental unmitigated result ($\lambda=1$) is $-1.793\pm0.004$ Ha, and the PIE mitigated result is $-1.86\pm0.01$ Ha. For the 12-qubit LiH molecule, the reference value is -8.875 Ha, the noise-free result is $-8.85\pm0.04$ Ha, the experimental unmitigated result is $-7.90\pm0.02$ Ha, and the PIE mitigated result is $-8.8\pm0.4$ Ha. 

\section{Discussion}
\noindent 
In this paper, we propose a linear runtime extrapolation-based error mitigation protocol built on the EMRE framework.
Unlike purely phenomenological data-fitting approaches, PIE's extrapolation fit is operationally guided from EMRE, without requiring explicit hardware noise characterization. 
This derivation singles out a specific point in the full ZNE parameter space and also provides a clear physical meaning to the parameters in the fit.
In this sense, PIE is not a simplified exponential ZNE, but a physically inspired specialization of ZNE where the extrapolation fit is derived rather than selected heuristically.
Furthermore, this operational grounding also explains why the parameters of heuristic ZNE approaches cannot be assigned physically meaningful interpretations or used for tasks such as circuit certification.

Through both numerical simulations and experimental demonstrations, we show that the protocol performs optimally in low-noise regimes where early FTQC algorithms are expected to operate or for short-depth circuits.
Moreover, PIE provides a clear physical interpretation to the experimental data fitting process and its associated parameters, a feature that is often absent in extrapolation-based mitigation techniques. 
Specifically, the slope of the fit approximates the max-relative entropy, a measure that quantifies the distance between the ideal and the noisy quantum circuit. 
This information enables users to assess the impact of noise on their circuits and determine whether the collected data remains meaningful.
In doing so, PIE not only extracts physically relevant quantities that characterize error mechanisms, but also facilitates a form of hardware certification where the slope serves as a proxy for measuring the deviation between implemented and ideal circuit behavior.
This dual capability distinguishes PIE from alternative mitigation methods such as conventional ZNE, which rely solely on empirical fitting without offering insight into the underlying error structure or hardware quality.
Since PIE’s extrapolation function is derived from the first principles under physical assumptions involving max-relative entropy, this level of interpretability cannot, in general, be obtained from heuristically motivated ZNE fits.

From an operational perspective, the PIE method offers efficient error mitigation with enhanced robustness and notably reduced variance compared to non-linear exponential fitting techniques.
This enhanced stability and variance stems from the linear extrapolation model enabled by the restricted evolution framework of EMRE. Furthermore, unlike recent learning-based methods, PIE does not require a pre-trained model or a large amount of data, while offering robust performance in the low-noise regime relevant to PIE compared to methods such as Clifford Data Regression (see Supp. Info. Sec. 4).


Looking ahead, a key direction is to extend this protocol to handle higher-noise settings and integrating it with fault-tolerance schemes to provide a scalable pathway toward reliable quantum computation. 
The challenge will be to design such methods to remain fully operationally meaningful rather than heuristic. 
Another direction for generalization is to consider the case of variable noise throughout the circuit.
To tackle variable noise, one approach can be to fold gates at their occurrence. In this way, one can have a higher confidence that the inverse unitary gate is not experiencing a different noise than the original unitary gate itself. Adaptive folding~\cite{Koenig2025May}, use of purity~\cite{jin_purity_2024} and other approaches~\cite{PhysRevA.110.042625, Hour2024Jan} were recently proposed for ZNE to tackle this practical problem; the question would be to incorporate it with a non-heuristically motivated extrapolation fit such as PIE.
Moreover, beyond full circuit folding, PIE may benefit from the use of more accurate noise amplification techniques, such as partial folding, analog noise strength control~\cite{Temme2017Nov}, or circuit quantum unoptimization~\cite{arxiv_2503.06341}. The combination of PIE with randomized compiling can also improve the performance of the method in the presence of coherent errors~\cite{Kurita_Synergetic_2022}.
It is also interesting to see how PIE can be used as an efficient computational tool to approximate max relative entropy in resource theoretic tasks.
Another open avenue of research is to develop efficient methods to compute the exact distance between the ideal and erroneous circuit, further strengthening the interpretability and diagnostic power of error mitigation.\\


\noindent \textbf{\large Data availability}\\
All relevant data were available in the main text and Supporting Information and can be obtained from the authors upon request. Source data are provided with this paper.\\

\noindent \textbf{\large Code availability}\\
All the software used in this work is open source. No specific software was developed for this work.

\bibliography{ref}

\newpage
\noindent \textbf{\large Acknowledgments}\\
We would like to thank our management executives - Kevin Ferreira, Yipeng Ji, Paria Nejat of LG Electronics Toronto AI Lab and Sean Kim of LG Electronics AI Lab, for their constant support throughout this work. P.D.-V. acknowledges support by CSIC iMOVE Program, Misiones Ciencia e Innovación Program (CUCO) under Grant MIG20211005, the European project PROMISCE, and the Alliance of Technology Centres for innovation in Quantum Computing applications for the enterprise ARQA (CER-20231031) funded by the CERVERA Research Programme of CDTI (Centre for Technological Development and Innovation). \\

\noindent \textbf{\large Author contributions}\\
P.D.-V., G.S., and T.H.K. conceived the project. G.S. developed the theoretical framework. P.D.-V. and G.S. designed the protocols.
P.D.-V., G.S., and T.H.K. analyzed the data.
P.D.-V. conducted the experiments on IBMQ hardware and performed the numerical simulations. 
J.H.L. helped with the addition of quantum chemistry example. J.B. and J.H.L. provided support with experimental and numerical simulation.
All authors discussed the results and contributed to the manuscript writing.\\

\noindent \textbf{\large Competing interests}\\
The authors declare no competing interests.\\


\appendix
\onecolumngrid


\makeatother

\newpage

\section*{Supplemental Information for ``Physically motivated extrapolation for quantum error mitigation"}
\section{Inverse EMRE technique}\label{sec:inverse_emre}
\setcounter{equation}{0}
In this section, we discuss another way to implement EMRE on a quantum computer. We call it the inverse EMRE technique inspired from the inverse method to implement PEC.
We also report numerical and experimental results obtained by using this technique and compare the performance of EMRE and HEMRE implemented using this way with that of PEC. We also mention the challenges in using this method and why the PIE method is preferred over this implementation method.
\subsection{Method}
Error mitigation by Restricted Evolution (EMRE) works by restricting the evolution of the input quantum state~\cite{SK2024emre}.
The output obtained from this restricted evolution gives the estimate of the ideal expectation value.
EMRE trades off sampling efficiency with a small non-zero bias.
The restriction is performed in such a way that sampling the circuit is efficient.
In Ref.~\cite{SK2024emre}, two of the authors of this paper introduced the concept of a generalized quasi-probabilistic decomposition (GQPD) for a unitary channel $\mU$. This decomposition is given by
\eqs{\mU = s\mB - (s-1)\mN\,,}
where $\mB$ is a convex combination of implementable operations and $\mN$ is some quantum channel.
By restricting the evolution of the input state to $\mB$, that is, by ignoring the negative part of the above decomposition, the authors showed that error mitigated results can be obtained with an efficient constant sampling overhead.

The above protocol relies on obtaining a generalized QPD for each gate in the circuit.
The smallest bias is achieved when the generalized QPD used is optimal. In~\cite{SK2024emre}, it has been shown that finding the optimal generalized QPD can be cast as an optimization problem. 
This optimization problem requires a characterization of the noise in the hardware to characterize the set of noisy operations, making it challenging to obtain the optimal solution.
Thus, with unknown hardware noise, it seems difficult to practically implement and mitigate noise using EMRE.

To overcome the above implementation challenge, inspired from the inverse method to perform PEC, we develop the inverse EMRE method.
Let $G$ denote an ideal gate to be implemented in a circuit. 
After noise acts on the gate, the erroneous gate $G^{\rm err}$ can be written as $G^{\rm err} = \mE\circ G$ where $\mE$ represents the noise acting after the gate.
In other words, the ideal gate can be expressed as $G = \mE^{-1}\circ G^{\rm err}$.
Thus, in order to apply the ideal gate, we need to apply an extra channel that is close to $\mE^{-1}$ after the erroneous gate $G^{\rm err}$ to reduce the effects of noise.

To apply $\mE^{-1}$ experimentally and perform error mitigation, we first need to characterize the noise. 
We assume that the noise acting on the system is a random Pauli noise. 
Using gate set tomography, we can construct the Pauli transfer matrix (PTM) for the erroneous gates.
Denoting the PTM of the ideal gate, the erroneous gate and the noise as $R^G, R^{G^{\rm err}},$ and $R^{\mE}$, respectively, we can represent the PTM of noise $\mE$ as $R^{\mE} = R^{G^{\rm err}}\circ {R^{G}}^{-1}$.
By experimentally computing the PTM of the erroneous gate $R^{G^{\rm err}}$, we can compute the PTM of noise using the given equation (since the PTM of the ideal gate is assumed to be known in theory).
Next, by keeping Paulis as the set of implementable operations, we can find a generalized QPD of ${R^{\mE}}^{-1}$.
The generalized QPD of ${R^{\mE}}^{-1}$ will have the following form:
\eqs{
{R^{\mE}}^{-1} = s\sum_i p_i R^{P_i}- (s-1)R^{\mM}\,,
\label{eq:invertechniquewhole}}
where $p_i$ represents the probabilities of the occurrence of Pauli $P_i$.
To perform EMRE, we can restrict the above decomposition to the positive part and approximate the ideal gate as follows:
\eqs{G \approx s\left(\sum_i p_i P_i\right)\circ G^{\rm err}\,.\label{eq:inverseEMREaverage}}
Practically, the above equation implies that we sample the Paulis from the probability distribution and apply them after the erroneous gate instance and multiply the final result by $s$.

\subsection{Numerical analysis}

\begin{figure*}
\includegraphics[width=1\linewidth]{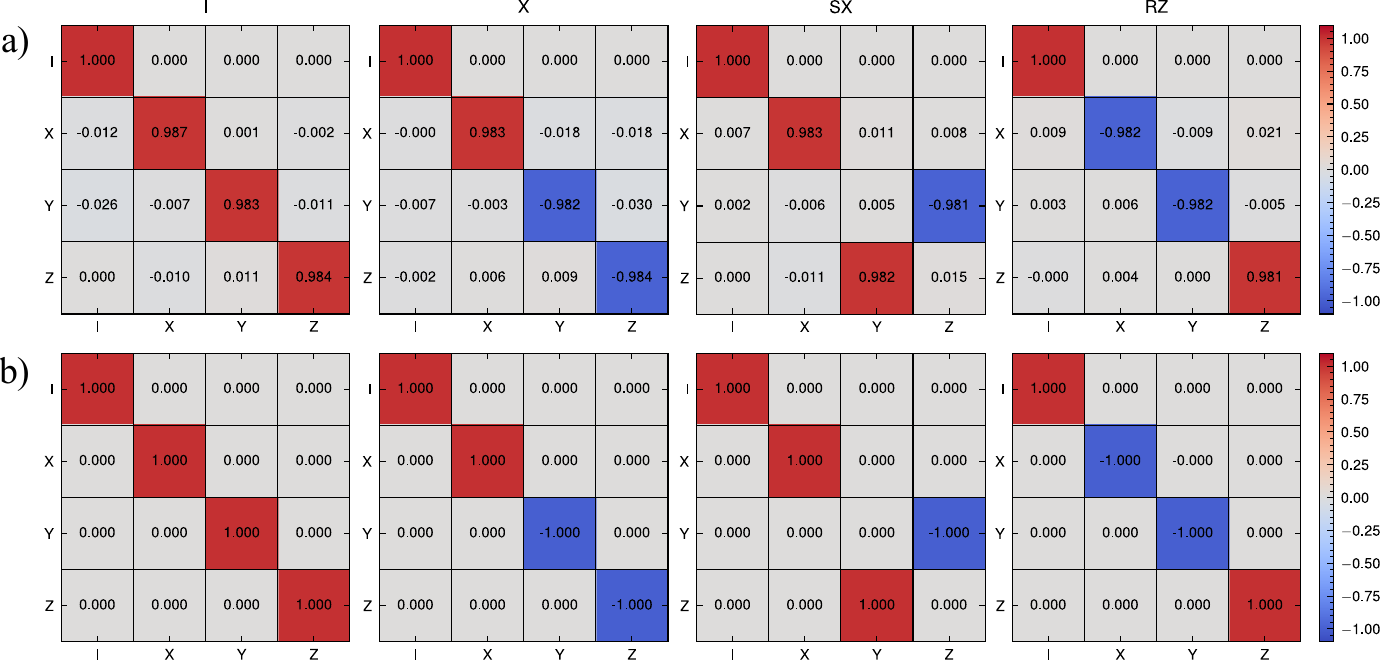}
\caption{\label{fig_ptms_singlequbit} (Color online) Examples of the Pauli transfer matrices representation (PTMs) of the identity (I), Pauli-X (X), Sqrt-X (SX), and rotation of angle $\pi$ around the Z axis (RZ) single-qubit gates. (a) Erroneous gate PTMs obtained from numerical simulations based on a noise model provided by qiskit that mimics the estimated noise of the IBM Eagle superconducting processor ibm kyiv using 4096 shots, (b) Ideal gate PTMs obtained from noise-free statevector simulator.}
\end{figure*}

\begin{figure*}
\includegraphics[width=0.96\linewidth]{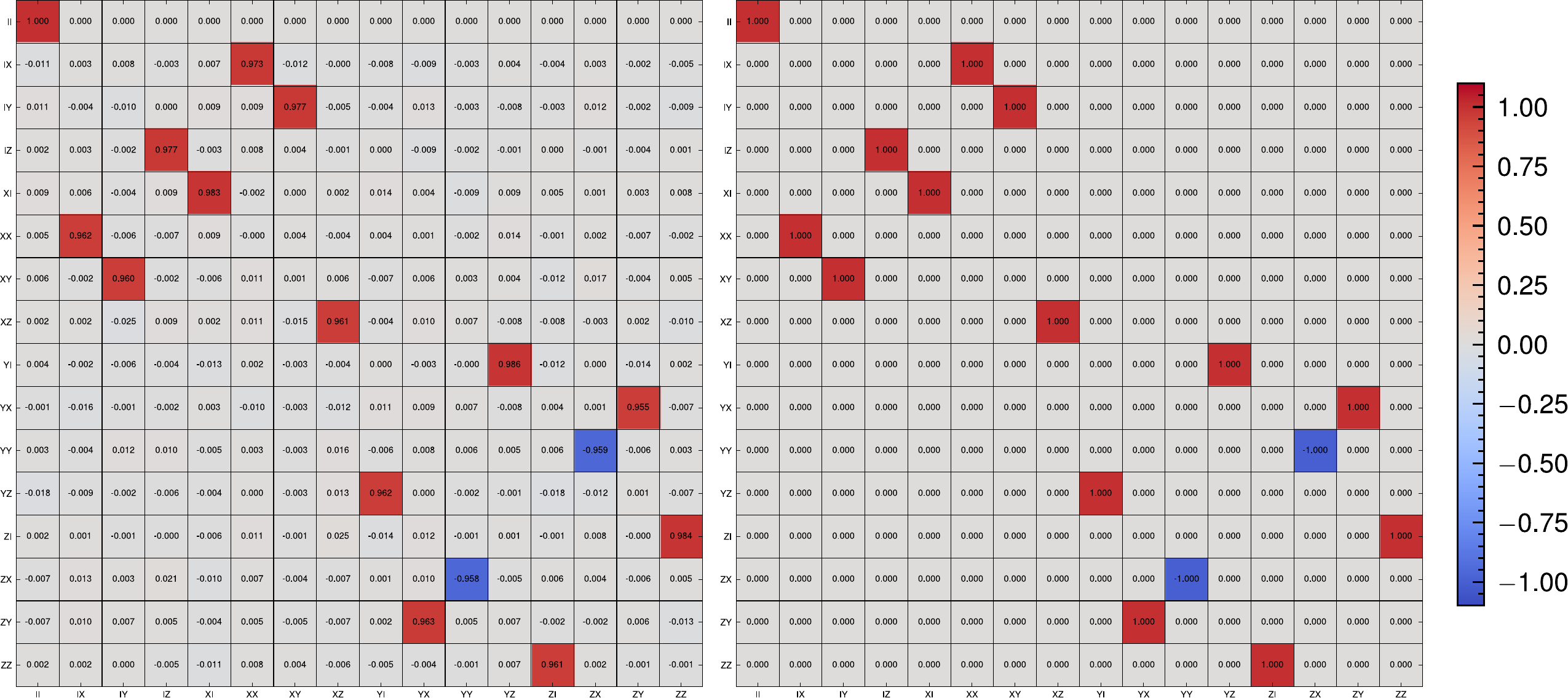}
\caption{\label{fig_ptms_twoqubit} (Color online) Examples of the Pauli transfer matrices representation (PTMs) of the CNOT gate. (left) Erroneous gate PTM obtained from numerical simulations based on a noise model provided by qiskit that mimics the estimated noise of the IBM Eagle superconducting processor ibm kyiv using 4096 shots, (right) Ideal gate PTM obtained from noise-free statevector simulator.}
\end{figure*}

\begin{figure*}
\includegraphics[width=1.0\linewidth]{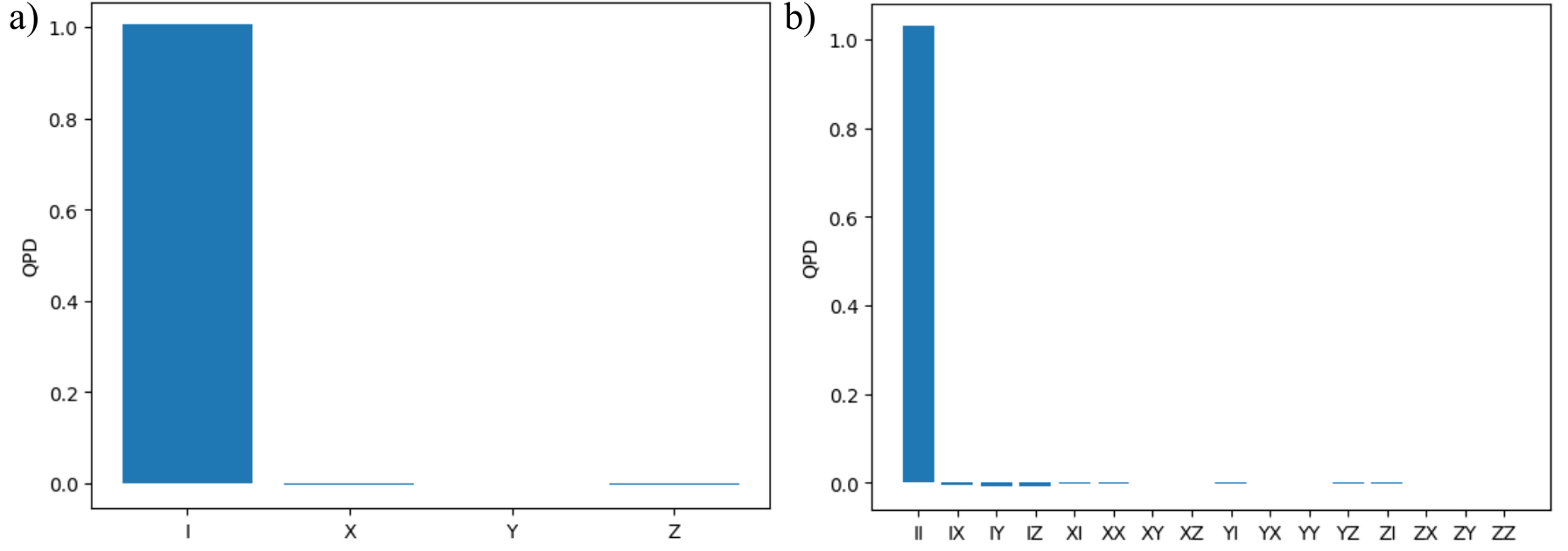}
\caption{\label{fig_qpds} (Color online) Quasi-probabilistic decompositions (QPDs) of the inverse noise operations assuming random Pauli noise of (a) the experimental single-qubit gate X, (b) the experimental two-qubit gate CNOT. The QPDS were obtained from numerical simulations based on a noise model provided by qiskit that mimics the estimated noise of the IBM Eagle superconducting processor ibm kyiv using 4096 shots.}
\end{figure*}

In the first place, the PTM representation $R^{\mE^{-1}}$ of the inverse noise channel $\mE^{-1}$ is estimated for each implemented single- and two-qubit gate.
This is done by computation of the PTM representation of the ideal $G$ and erroneous $G^{\rm err}$ such that ${R^{\mE}}^{-1} = R^{G}\circ {R^{G^{\rm err}}}^{-1}$, where
\begin{equation}
    R^{\Lambda} = \dfrac{1}{2^{N}} \tr\left[P_i\Lambda(P_j)\right]
    \label{eq_PTM}
\end{equation}
with $\Lambda$ being the one- ($N=1$) or two-qubit ($N=2$) quantum channels ${\mE}^{-1}$, $G$ or $G^{\rm err}$, and $P_{i}$ are the elements is the $N$-qubit Pauli basis ($\{I,X,Y,Z\}$ for $N=1$ and $\{II,IX,IY,IZ,XI,XX,XY,XZ,YI,YX,YY,YZ,ZI,ZX,ZY,ZZ\}$ for $N=2$). 
In~\figref{fig_ptms_singlequbit} and~\figref{fig_ptms_twoqubit} we show some examples of erroneous gate PTMs, $R^{G^{\rm err}}$, and ideal gate PTMs, $R^{G}$, obtained from numerical simulations and Eq.~\eqref{eq_PTM}. 
Assuming random Pauli noise, the PTM representation of the inverse noise channel ${R^{\mE}}^{-1}$ together with the PTM representation of erroneous Pauli gates (where $\Lambda=P_i$) allows for the calculation of a GQPD $\{q^G_i\}$ by solving an optimization problem as given in Ref.~\cite{SK2024emre}, so that ${R^{\mE}}^{-1} \approx \sum_{i} q^G_i P_i$ and $\sum_iq^G_i=1$.
Recall that a QPD is also a GQPD.
In~\figref{fig_qpds} we show examples of QPDs of the inverse noise operations associated with the single-qubit X gate and the two-qubit controlled-NOT gate in numerical simulations. 

Now, to implement the inverse EMRE technique, we restrict the QPD to its positive part.
The normalized coefficients of the positive part represent the probabilities $\{p^G_i=q^G_i/s^G | q^G_i>0\}_i$ of the occurrence of Pauli gates $\{P_i\}_i$ such that each ideal quantum gate can be approximated by sampling from the $N$-qubit Pauli basis $\{P_i\}$ with probabilities $p^G_i$ as depicted in Eq.~\eqref{eq:inverseEMREaverage}. 
The robustness $s^{G}$ of the gate $G$ is the sum of positive coefficients of the QPD, i.e., $s=\sum_iq^G_i$. 
In practice, a number of circuits $K$ is sampled where the gates are probabilistically chosen following Eq.~\eqref{eq:inverseEMREaverage}, the expectation value of the observable of interest is computed for each sampled circuit $\langle\mO\rangle_{k}$ with $k\in[1,2,...,K]$, and the mitigated estimate is calculated as the average of those expectation values,
\begin{equation}
    \langle\mO\rangle_{\textnormal{mitigated}} = \frac{s}{K}\sum_{k=1}^K \langle\mO\rangle_{k}\;\;,
\end{equation}
where $s = \prod_G s^G$. The inverse technique to perform Probabilistic Error Cancellation (PEC) follows similarly with $\{p^{(PEC)G}_i\}=\{|q^G_i|\}$ and $s^{(PEC)G}=\sum_ip^{(PEC)G}_i$. Contrary to EMRE, PEC provides an unbiased result provided a sufficiently large number of circuit samples, however, it has been shown that EMRE achieves more accurate estimates in a scenario with reduced number of samples. 
In~\cite{SK2024emre}, the Hybrid Error Mitigation By Restricted Evolution (HEMRE) is also introduced, where only a subset of gates is restricted to positive QPDs. In particular, we consider an HEMRE protocol where $\{p^{(HEMRE)G}_i\}=\{q^G_i | q^G_i>0\}$ if $G$ is a single-qubit gate, $\{p^{(HEMRE)G}_i\}=\{|q^G_i|\}$ if $G$ is a two-qubit gate, and $s^{(HEMRE)G}=\sum_ip^{(HEMRE)G}_i$.

\begin{figure*}
\includegraphics[width=1\linewidth]{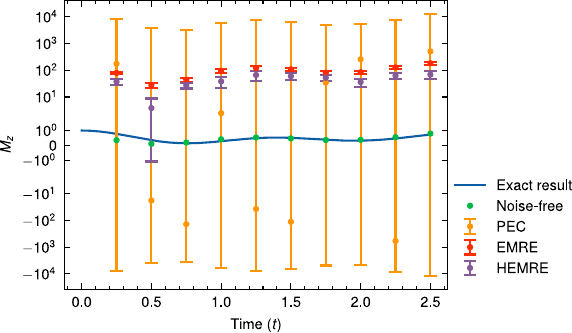}
\caption{\label{fig_PECvsEMREvsHEMRE_simulation} (Color online) Numerical results obtained using the inverse technique described in Sec.~\ref{sec:inverse_emre} for PEC, EMRE, and HEMRE mitigation protocols. The results were obtained for two Trotter steps simulation of the one-dimensional Ising chain with $N = 6$ spins at different times using a noise model provided by qiskit that mimics the estimated noise of the IBM Eagle superconducting processor ibm kyiv using 4096 shots. We show the mitigated result (dots) calculated from the average of $K=1200$ samples together with standard deviations (bars). We also display the noise-free result from the quantum circuit execution as well the exact result of the one-dimensional Ising chain evolution without the associated Trotter error.}
\end{figure*}

\begin{figure*}
\includegraphics[width=1.0\linewidth]{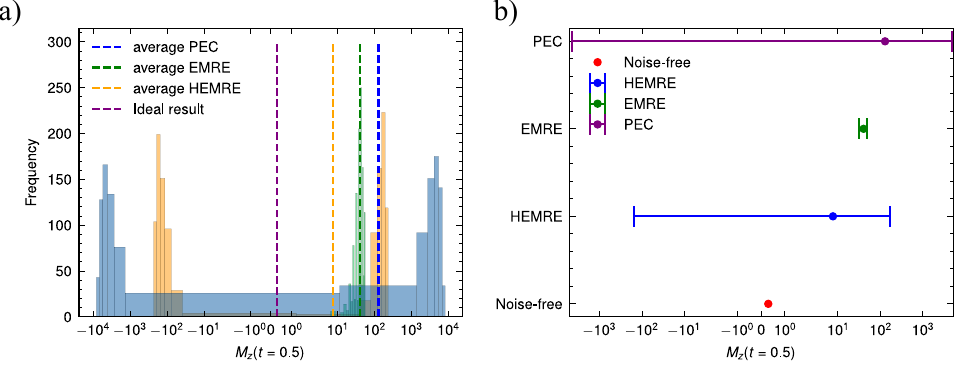}
\caption{\label{fig_PECvsEMREvsHEMRE} (Color online) Experimental results obtained using the inverse technique described in Sec.~\ref{sec:inverse_emre} for PEC, EMRE, and HEMRE mitigation protocols. The results were obtained for one Trotter step simulation of the one-dimensional Ising chain with $N = 20$ spins at time $t=0.5$ on the IBM Heron superconducting processor ibm torino using 4096 shots. (a) Frequency distribution of global magnetization 1200 estimates sampled from the Paulis distribution and the erroneous gates whose average provides a error-mitigated biased result (see Eq.~\eqref{eq:invertechniquewhole} and ~\eqref{eq:inverseEMREaverage}). (b) Averages (dots) together with standard deviations (bars).}
\end{figure*}

As done in the manuscript, we tested the previous mitigation protocols on the time dynamics simulation of the one-dimensional quantum Ising spin chain Hamiltonian with periodic boundary condition ($\hbar=1$ throughout),
\begin{equation}
    \hat{H} = -\sum_{i=0}^{N-2} \sigma^x_{i}\sigma^x_{i+1} - \sigma^x_{N-1}\sigma^x_{0} -  \Gamma \sum_{i=0}^{N-1} \sigma^z_{i} \,,
    \label{eq_Isingchain_supp}
\end{equation}
with $N$ the number of qubits, and $\Gamma=0.5$ the transverse field strength. The dynamics is approximated via the first-order Trotter-Suzuki decomposition using one Trotter step to compute the global transverse magnetization at time $t=0.5$. We test the three protocols using the described inverse technique with a constant number of sampled circuits $K=1200$. In Figure~\ref{fig_PECvsEMREvsHEMRE_simulation} and~\ref{fig_PECvsEMREvsHEMRE} we show the numerical and experimental results respectively. Note that the time between the QPD calculation and the sampling must be as small as possible to reduce errors coming from physical fluctuations occurring in the quantum computer. 

We observe that this technique leads to approximation errors, notable computational overhead due to QPD or GQPD calculation and circuit sampling, and a large bias arising from the large value of robustness $s$, which motivates the use of more efficient implementations of EMRE such as the physics-inspired extrapolation (PIE) method presented in the manuscript.


\newpage
\hspace{0cm}

\section{Numerical simulation results of PIE under realistic Noise Models}
\label{sec_noisemodels}

\begin{figure*}
\includegraphics[width=1.0\linewidth]{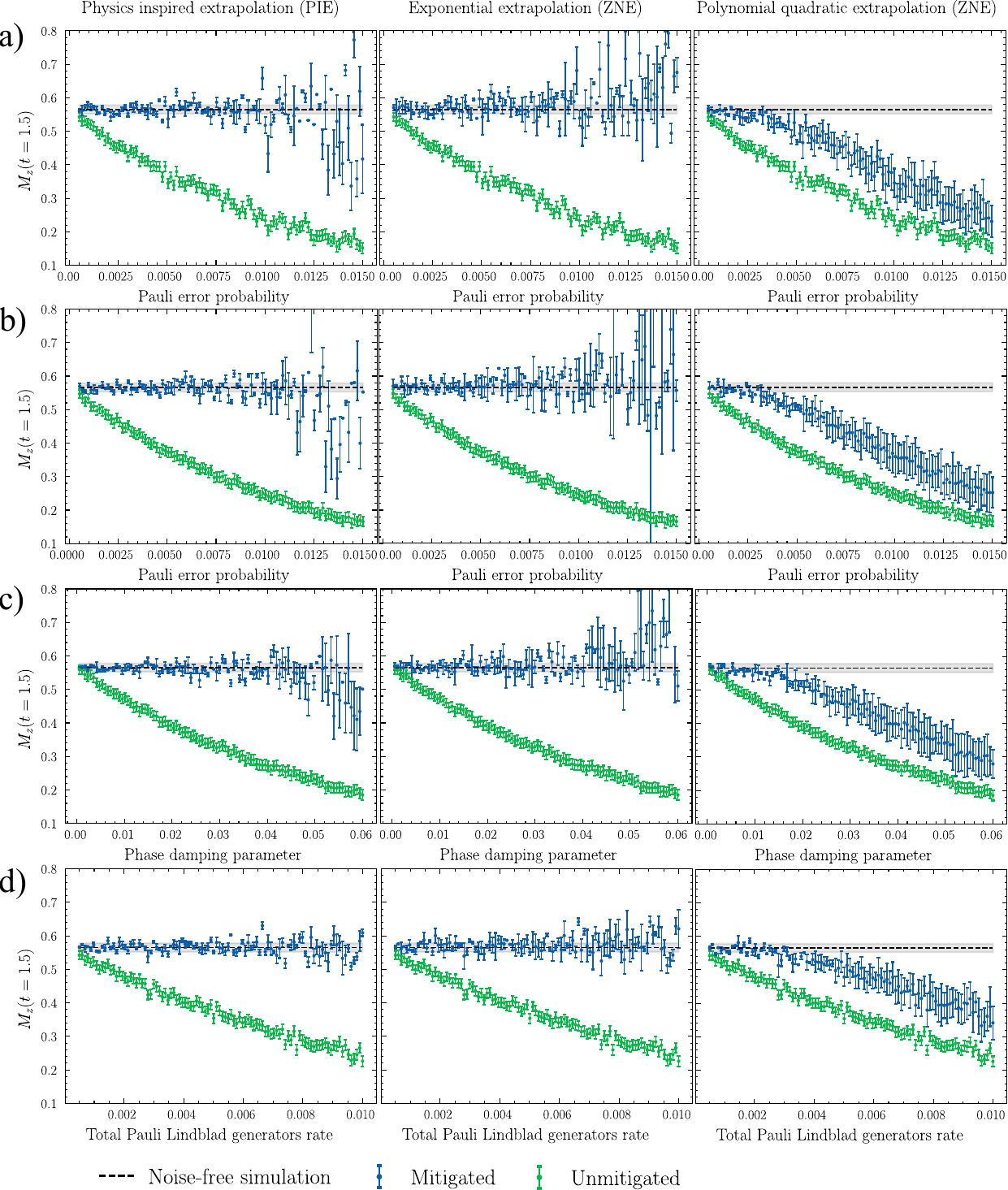}
\caption{\label{fig_allnoisemodels} (Color online) Numerical results of the global magnetization $M_z$ on the three Trotter steps simulation of the
one-dimensional Ising chain at time $t=1.5$ with N = 8 spins using the noise models expanded in section~\ref{sec_noisemodels}, namely, mixed Pauli quantum error channel based on (a) fully random probabilities and (b) only incremental probabilities, (c) phase damping quantum error channel, and (d) Pauli quantum error channel generated by Pauli Lindblad dissipators. We display the error mitigated results (dots) from PIE, exponential ZNE, and polynomial quadratic ZNE techniques, together with the standard deviation of the results (bars). We also show the unmitigated or noisy result and the noise-free simulation, with the corresponding standard deviation (bars and shaded area, respectively). All experiments were performed using four extrapolation points and 4096 shots.}
\end{figure*}

In this section, we further test the performance of our PIE method under different practically relevant noise models using the same setup as mentioned in the main text, i.e., using the time dynamics simulation of the one-dimensional quantum Ising spin chain Hamiltonian (see description around Eq.~(1) in the main text).
Specifically, we consider the following noise models: inhomogeneous Pauli noise, partially depolarizing noise, partially dephasing noise, and the Pauli Lindblad noise.
The results under the partially depolarizing noise are presented in the main text.
Here, in~\figref{fig_allnoisemodels} we present the results for the other noise models.
The results in~\figref{fig_allnoisemodels}(a) are obtained using random Pauli noise probabilities where the total noise probability increases but the noise probability of occurrence of the X-, Y-, and Z-noise is completely random.
We report this data in Table \ref{tab:pauli-probabilities}.
In~\figref{fig_allnoisemodels}(b), the results are obtained by incrementing the X-, Y-, and Z-noise probabilities as the total noise probability increases.
We report this data in Table~\ref{tab:pauli-probabilities_2}.
The results in~\figref{fig_allnoisemodels}(c) and~\figref{fig_allnoisemodels}(d) correspond to the case of partially dephasing and Pauli Lindblad noise (where the rates for the Pauli generators are obtained as the random Pauli noise model of~\figref{fig_allnoisemodels}(a) and reported in Table~\ref{tab:pauliLinbland-probabilities}), respectively. 
Note that for all the noise models analyzed, when the noise probability is very high, there is no meaningful quantum information that can be recovered from the quantum circuit measurements and the noise-amplified numerical expectation values tend to fluctuate around zero. These fluctuations can result in a mixture of positive and negative values. To avoid singularities caused by the logarithm in numerical simulations, we set a value of $10^{-6}$ to any expectation value less than or equal to zero, which in practice does not affect the mitigation performance.

\begin{table}
\centering
\begin{minipage}{0.5\textwidth}
\centering
\begin{tabular}{cccc}
\hline
$p_X$ & $p_Y$ & $p_Z$ & Total Error Probability \\
\hline
0.00024 & 8e-05 & 0.00018 & 0.0005 \\
8e-05 & 0.00035 & 0.00022 & 0.00065 \\
0.00025 & 0.00018 & 0.00036 & 0.00079 \\
0.00023 & 0.00043 & 0.00029 & 0.00095 \\
0.00046 & 0.00015 & 0.00047 & 0.00108 \\
0.00075 & 0.0001 & 0.00038 & 0.00123 \\
0.00036 & 0.00047 & 0.00055 & 0.00138 \\
0.00064 & 0.00043 & 0.00045 & 0.00152 \\
0.00068 & 0.00042 & 0.00057 & 0.00167 \\
0.00019 & 0.00109 & 0.00054 & 0.00182 \\
0.00159 & 0.00035 & 2e-05 & 0.00196 \\
0.00023 & 0.00085 & 0.00103 & 0.00211 \\
5e-05 & 0.00139 & 0.00082 & 0.00226 \\
0.00028 & 0.00104 & 0.00108 & 0.0024 \\
0.00148 & 4e-05 & 0.00103 & 0.00255 \\
0.00049 & 0.00075 & 0.00145 & 0.00269 \\
0.00097 & 0.00141 & 0.00046 & 0.00284 \\
0.00011 & 0.00173 & 0.00116 & 0.003 \\
0.00073 & 0.00086 & 0.00155 & 0.00314 \\
0.00167 & 0.001 & 0.00062 & 0.00329 \\
0.00138 & 0.00047 & 0.00158 & 0.00343 \\
0.00099 & 0.00185 & 0.00073 & 0.00357 \\
0.00157 & 0.00102 & 0.00114 & 0.00373 \\
0.00137 & 0.00081 & 0.00169 & 0.00387 \\
0.00065 & 0.00059 & 0.00278 & 0.00402 \\
0.00151 & 0.00188 & 0.00077 & 0.00416 \\
0.00334 & 0.00051 & 0.00046 & 0.00431 \\
0.00213 & 0.0007 & 0.00163 & 0.00446 \\
0.00199 & 0.00155 & 0.00105 & 0.00459 \\
0.00032 & 0.00223 & 0.0022 & 0.00475 \\
0.00078 & 0.00076 & 0.00335 & 0.00489 \\
0.00019 & 0.00345 & 0.0014 & 0.00504 \\
0.00257 & 0.00185 & 0.00077 & 0.00519 \\
0.00017 & 0.00408 & 0.00108 & 0.00533 \\
0.00231 & 0.00136 & 0.00181 & 0.00548 \\
0.00031 & 0.00309 & 0.00223 & 0.00563 \\
0.00162 & 0.00134 & 0.00281 & 0.00577 \\
0.00069 & 0.00273 & 0.0025 & 0.00592 \\
0.00249 & 0.00126 & 0.00231 & 0.00606 \\
0.00048 & 0.00074 & 0.00499 & 0.00621 \\
0.00323 & 0.00073 & 0.0024 & 0.00636 \\
0.00358 & 0.00149 & 0.00144 & 0.00651 \\
0.00481 & 0.00012 & 0.00173 & 0.00666 \\
0.00386 & 0.00037 & 0.00257 & 0.0068 \\
0.00256 & 0.00068 & 0.0037 & 0.00694 \\
0.0024 & 0.0022 & 0.00249 & 0.00709 \\
0.00327 & 6e-05 & 0.0039 & 0.00723 \\
0.0008 & 0.00076 & 0.00582 & 0.00738 \\
0.00665 & 0.00011 & 0.00077 & 0.00753 \\
0.00089 & 0.00164 & 0.00515 & 0.00768 \\
\hline
\end{tabular}
\end{minipage}%
\hfill
\begin{minipage}{0.5\textwidth}
\centering
\begin{tabular}{cccc}
\hline
$p_X$ & $p_Y$ & $p_Z$ & Total Error Probability \\
\hline
0.00092 & 0.00489 & 0.00201 & 0.00782 \\
0.00327 & 0.00095 & 0.00375 & 0.00797 \\
0.00155 & 0.00047 & 0.00609 & 0.00811 \\
0.00313 & 0.00275 & 0.00238 & 0.00826 \\
0.00235 & 0.00319 & 0.00286 & 0.0084 \\
0.00385 & 0.00395 & 0.00076 & 0.00856 \\
0.00054 & 0.00472 & 0.00344 & 0.0087 \\
0.00312 & 0.00301 & 0.00272 & 0.00885 \\
0.00332 & 0.00338 & 0.00229 & 0.00899 \\
0.00116 & 0.00051 & 0.00747 & 0.00914 \\
0.00267 & 0.00045 & 0.00616 & 0.00928 \\
0.00394 & 0.00292 & 0.00258 & 0.00944 \\
0.00049 & 0.00734 & 0.00174 & 0.00957 \\
0.00313 & 0.00377 & 0.00283 & 0.00973 \\
0.00601 & 0.00264 & 0.00122 & 0.00987 \\
0.00222 & 0.00486 & 0.00294 & 0.01002 \\
0.00187 & 0.00824 & 6e-05 & 0.01017 \\
0.00281 & 0.0066 & 0.00091 & 0.01032 \\
0.00498 & 0.00405 & 0.00143 & 0.01046 \\
0.00616 & 0.00436 & 9e-05 & 0.01061 \\
0.00446 & 0.00375 & 0.00255 & 0.01076 \\
0.00415 & 0.00192 & 0.00483 & 0.0109 \\
0.00378 & 0.00323 & 0.00404 & 0.01105 \\
0.00341 & 0.0074 & 0.00039 & 0.0112 \\
0.00413 & 0.0041 & 0.0031 & 0.01133 \\
0.00858 & 0.00087 & 0.00203 & 0.01148 \\
0.00715 & 0.00395 & 0.00053 & 0.01163 \\
0.00464 & 0.00204 & 0.00509 & 0.01177 \\
0.0052 & 0.00516 & 0.00157 & 0.01193 \\
0.00284 & 0.00038 & 0.00885 & 0.01207 \\
0.00225 & 0.00315 & 0.00683 & 0.01223 \\
0.00131 & 0.00498 & 0.00608 & 0.01237 \\
0.00275 & 0.00793 & 0.00182 & 0.0125 \\
0.00054 & 0.00712 & 0.005 & 0.01266 \\
0.00098 & 0.00529 & 0.00654 & 0.01281 \\
0.00251 & 0.00538 & 0.00506 & 0.01295 \\
0.0011 & 0.00563 & 0.00636 & 0.01309 \\
0.00341 & 0.00611 & 0.00372 & 0.01324 \\
0.00243 & 0.00847 & 0.00249 & 0.01339 \\
0.00183 & 0.00502 & 0.00668 & 0.01353 \\
0.00468 & 0.00823 & 0.00077 & 0.01368 \\
0.00615 & 0.00745 & 0.00022 & 0.01382 \\
0.00938 & 0.0029 & 0.0017 & 0.01398 \\
0.00422 & 0.00255 & 0.00735 & 0.01412 \\
0.00363 & 0.00446 & 0.00619 & 0.01428 \\
0.00259 & 0.00609 & 0.00573 & 0.01441 \\
0.00627 & 0.00158 & 0.00672 & 0.01457 \\
0.0074 & 0.00479 & 0.00252 & 0.01471 \\
0.00331 & 0.00767 & 0.00387 & 0.01485 \\
0.00263 & 0.00941 & 0.00296 & 0.015 \\
\hline
\end{tabular}
\end{minipage}
\caption{Probabilities of Pauli X, Y, and Z errors used on the simulated mixed Pauli quantum error channel across the 100 samples reported in Fig.~\ref{fig_allnoisemodels}a.}
\label{tab:pauli-probabilities}
\end{table}

\begin{table}
\centering
\begin{minipage}{0.5\textwidth}
\centering
\begin{tabular}{cccc}
\hline
$p_X$ & $p_Y$ & $p_Z$ & Total Error Probability \\
\hline
0.00017 & 0.00017 & 0.00017 & 0.00051 \\
0.00031 & 0.00017 & 0.00017 & 0.00065 \\
0.00036 & 0.00022 & 0.00022 & 0.0008 \\
0.00036 & 0.00026 & 0.00031 & 0.00093 \\
0.00041 & 0.00031 & 0.00036 & 0.00108 \\
0.00046 & 0.00036 & 0.00041 & 0.00123 \\
0.00051 & 0.00041 & 0.00046 & 0.00138 \\
0.00051 & 0.00051 & 0.00051 & 0.00153 \\
0.00065 & 0.00051 & 0.00051 & 0.00167 \\
0.0007 & 0.00056 & 0.00056 & 0.00182 \\
0.0008 & 0.00056 & 0.00061 & 0.00197 \\
0.00085 & 0.00056 & 0.0007 & 0.00211 \\
0.00085 & 0.00061 & 0.0008 & 0.00226 \\
0.0009 & 0.0007 & 0.0008 & 0.0024 \\
0.00095 & 0.0007 & 0.0009 & 0.00255 \\
0.00105 & 0.0007 & 0.00095 & 0.0027 \\
0.00105 & 0.0008 & 0.001 & 0.00285 \\
0.00105 & 0.0009 & 0.00105 & 0.003 \\
0.00109 & 0.00095 & 0.00109 & 0.00313 \\
0.00114 & 0.00105 & 0.00109 & 0.00328 \\
0.00114 & 0.00105 & 0.00124 & 0.00343 \\
0.00119 & 0.00105 & 0.00134 & 0.00358 \\
0.00124 & 0.00109 & 0.00139 & 0.00372 \\
0.00124 & 0.00119 & 0.00144 & 0.00387 \\
0.00129 & 0.00124 & 0.00148 & 0.00401 \\
0.00134 & 0.00129 & 0.00153 & 0.00416 \\
0.00134 & 0.00129 & 0.00168 & 0.00431 \\
0.00139 & 0.00134 & 0.00173 & 0.00446 \\
0.00139 & 0.00144 & 0.00178 & 0.00461 \\
0.00144 & 0.00153 & 0.00178 & 0.00475 \\
0.00148 & 0.00163 & 0.00178 & 0.00489 \\
0.00153 & 0.00168 & 0.00183 & 0.00504 \\
0.00158 & 0.00168 & 0.00192 & 0.00518 \\
0.00158 & 0.00168 & 0.00207 & 0.00533 \\
0.00158 & 0.00183 & 0.00207 & 0.00548 \\
0.00163 & 0.00183 & 0.00217 & 0.00563 \\
0.00163 & 0.00188 & 0.00227 & 0.00578 \\
0.00163 & 0.00197 & 0.00231 & 0.00591 \\
0.00168 & 0.00207 & 0.00231 & 0.00606 \\
0.00173 & 0.00207 & 0.00241 & 0.00621 \\
0.00173 & 0.00212 & 0.00251 & 0.00636 \\
0.00178 & 0.00222 & 0.00251 & 0.00651 \\
0.00183 & 0.00231 & 0.00251 & 0.00665 \\
0.00183 & 0.00246 & 0.00251 & 0.0068 \\
0.00183 & 0.00251 & 0.00261 & 0.00695 \\
0.00183 & 0.00261 & 0.00266 & 0.0071 \\
0.00188 & 0.00261 & 0.00275 & 0.00724 \\
0.00192 & 0.00266 & 0.0028 & 0.00738 \\
0.00192 & 0.00271 & 0.0029 & 0.00753 \\
0.00202 & 0.00275 & 0.0029 & 0.00767 \\
\hline
\end{tabular}
\end{minipage}%
\hfill
\begin{minipage}{0.5\textwidth}
\centering
\begin{tabular}{cccc}
\hline
$p_X$ & $p_Y$ & $p_Z$ & Total Error Probability \\
\hline
0.00207 & 0.0028 & 0.00295 & 0.00782 \\
0.00207 & 0.0029 & 0.003 & 0.00797 \\
0.00212 & 0.00295 & 0.00305 & 0.00812 \\
0.00217 & 0.00295 & 0.00314 & 0.00826 \\
0.00217 & 0.003 & 0.00324 & 0.00841 \\
0.00227 & 0.003 & 0.00329 & 0.00856 \\
0.00231 & 0.0031 & 0.00329 & 0.0087 \\
0.00236 & 0.0031 & 0.00339 & 0.00885 \\
0.00241 & 0.00314 & 0.00344 & 0.00899 \\
0.00251 & 0.00319 & 0.00344 & 0.00914 \\
0.00256 & 0.00329 & 0.00344 & 0.00929 \\
0.00261 & 0.00329 & 0.00354 & 0.00944 \\
0.00275 & 0.00329 & 0.00354 & 0.00958 \\
0.00275 & 0.00339 & 0.00358 & 0.00972 \\
0.00275 & 0.00349 & 0.00363 & 0.00987 \\
0.0029 & 0.00349 & 0.00363 & 0.01002 \\
0.0029 & 0.00354 & 0.00373 & 0.01017 \\
0.0029 & 0.00358 & 0.00383 & 0.01031 \\
0.0029 & 0.00363 & 0.00393 & 0.01046 \\
0.003 & 0.00368 & 0.00393 & 0.01061 \\
0.00305 & 0.00373 & 0.00397 & 0.01075 \\
0.0031 & 0.00378 & 0.00402 & 0.0109 \\
0.00314 & 0.00388 & 0.00402 & 0.01104 \\
0.00319 & 0.00397 & 0.00402 & 0.01118 \\
0.00319 & 0.00407 & 0.00407 & 0.01133 \\
0.00319 & 0.00417 & 0.00412 & 0.01148 \\
0.00319 & 0.00427 & 0.00417 & 0.01163 \\
0.00324 & 0.00437 & 0.00417 & 0.01178 \\
0.00329 & 0.00437 & 0.00427 & 0.01193 \\
0.00344 & 0.00437 & 0.00427 & 0.01208 \\
0.00354 & 0.00441 & 0.00427 & 0.01222 \\
0.00358 & 0.00451 & 0.00427 & 0.01236 \\
0.00358 & 0.00456 & 0.00437 & 0.01251 \\
0.00358 & 0.00461 & 0.00446 & 0.01265 \\
0.00358 & 0.00466 & 0.00456 & 0.0128 \\
0.00363 & 0.00471 & 0.00461 & 0.01295 \\
0.00373 & 0.00471 & 0.00466 & 0.0131 \\
0.00373 & 0.00476 & 0.00476 & 0.01325 \\
0.00378 & 0.0048 & 0.0048 & 0.01338 \\
0.00393 & 0.0048 & 0.0048 & 0.01353 \\
0.00393 & 0.0049 & 0.00485 & 0.01368 \\
0.00397 & 0.00495 & 0.0049 & 0.01382 \\
0.00402 & 0.005 & 0.00495 & 0.01397 \\
0.00402 & 0.00505 & 0.00505 & 0.01412 \\
0.00407 & 0.0051 & 0.0051 & 0.01427 \\
0.00417 & 0.00515 & 0.0051 & 0.01442 \\
0.00422 & 0.00515 & 0.0052 & 0.01457 \\
0.00422 & 0.00529 & 0.0052 & 0.01471 \\
0.00432 & 0.00534 & 0.0052 & 0.01486 \\
0.00432 & 0.00539 & 0.00529 & 0.015 \\
\hline
\end{tabular}
\end{minipage}
\caption{Probabilities of Pauli X, Y, and Z errors used on the simulated mixed Pauli quantum error channel across the 100 samples reported in Fig.~\ref{fig_allnoisemodels}b.}
\label{tab:pauli-probabilities_2}
\end{table}

\begin{table}
\centering
\begin{minipage}{0.5\textwidth}
\centering
\begin{tabular}{cccc}
\hline
$p_X$ & $p_Y$ & $p_Z$ & Total Error Probability \\
\hline
0.00033 & 0.00013 & 4e-05 & 0.0005 \\
0.00012 & 0.00031 & 0.00017 & 0.0006 \\
0.00032 & 0.00011 & 0.00027 & 0.0007 \\
0.00043 & 6e-05 & 0.0003 & 0.00079 \\
0.00038 & 0.00021 & 0.00029 & 0.00088 \\
0.00034 & 0.00038 & 0.00026 & 0.00098 \\
0.00053 & 0.00011 & 0.00044 & 0.00108 \\
0.00051 & 0.00062 & 4e-05 & 0.00117 \\
0.00064 & 0.00023 & 0.00039 & 0.00126 \\
0.00018 & 0.00055 & 0.00063 & 0.00136 \\
0.00064 & 0.00044 & 0.00039 & 0.00147 \\
0.00026 & 0.00038 & 0.00092 & 0.00156 \\
0.00092 & 0.00038 & 0.00035 & 0.00165 \\
0.00037 & 0.00065 & 0.00073 & 0.00175 \\
0.00098 & 0.00086 & 1e-05 & 0.00185 \\
0.00041 & 0.00088 & 0.00065 & 0.00194 \\
0.00061 & 0.00103 & 0.0004 & 0.00204 \\
0.00078 & 0.00083 & 0.00052 & 0.00213 \\
4e-05 & 0.00049 & 0.0017 & 0.00223 \\
0.00106 & 0.00059 & 0.00067 & 0.00232 \\
0.00034 & 0.00074 & 0.00134 & 0.00242 \\
0.00141 & 0.00014 & 0.00097 & 0.00252 \\
0.00062 & 0.00099 & 0.001 & 0.00261 \\
0.00151 & 0.00018 & 0.00102 & 0.00271 \\
0.00061 & 0.00198 & 0.00021 & 0.0028 \\
0.00029 & 0.00227 & 0.00034 & 0.0029 \\
0.00042 & 0.00055 & 0.00202 & 0.00299 \\
0.00197 & 4e-05 & 0.00108 & 0.00309 \\
0.00084 & 0.00067 & 0.00168 & 0.00319 \\
0.00152 & 0.00134 & 0.00043 & 0.00329 \\
0.00108 & 0.00083 & 0.00146 & 0.00337 \\
0.00192 & 0.00123 & 0.00033 & 0.00348 \\
0.00035 & 0.00092 & 0.0023 & 0.00357 \\
0.00037 & 0.00066 & 0.00264 & 0.00367 \\
0.00039 & 0.00183 & 0.00154 & 0.00376 \\
0.00302 & 0.0007 & 0.00014 & 0.00386 \\
0.00091 & 0.00266 & 0.00039 & 0.00396 \\
0.00145 & 0.00208 & 0.00052 & 0.00405 \\
0.00098 & 0.00145 & 0.00172 & 0.00415 \\
0.0013 & 0.00267 & 0.00027 & 0.00424 \\
0.00216 & 0.0016 & 0.00058 & 0.00434 \\
0.00053 & 0.00053 & 0.00337 & 0.00443 \\
0.00112 & 0.00025 & 0.00317 & 0.00454 \\
0.00217 & 0.0011 & 0.00136 & 0.00463 \\
0.00072 & 0.00205 & 0.00196 & 0.00473 \\
0.00037 & 0.00147 & 0.00298 & 0.00482 \\
0.00185 & 0.00086 & 0.00221 & 0.00492 \\
0.00105 & 0.00324 & 0.00072 & 0.00501 \\
0.00134 & 0.00081 & 0.00295 & 0.0051 \\
0.00259 & 0.00045 & 0.00216 & 0.0052 \\
\hline
\end{tabular}
\end{minipage}%
\hfill
\begin{minipage}{0.5\textwidth}
\centering
\begin{tabular}{cccc}
\hline
$p_X$ & $p_Y$ & $p_Z$ & Total Error Probability \\
\hline
0.00282 & 0.00128 & 0.0012 & 0.0053 \\
0.00352 & 0.00094 & 0.00094 & 0.0054 \\
0.00137 & 0.00253 & 0.00158 & 0.00548 \\
0.00175 & 0.00189 & 0.00194 & 0.00558 \\
8e-05 & 0.0023 & 0.0033 & 0.00568 \\
0.00056 & 0.00313 & 0.00209 & 0.00578 \\
0.00261 & 0.00085 & 0.00242 & 0.00588 \\
0.00106 & 0.00275 & 0.00216 & 0.00597 \\
0.00511 & 0.00044 & 0.00052 & 0.00607 \\
0.00208 & 0.00185 & 0.00223 & 0.00616 \\
0.00233 & 0.00101 & 0.00292 & 0.00626 \\
0.00105 & 0.00243 & 0.00287 & 0.00635 \\
0.00359 & 0.00166 & 0.0012 & 0.00645 \\
0.00436 & 6e-05 & 0.00212 & 0.00654 \\
0.00078 & 0.00315 & 0.00271 & 0.00664 \\
0.00248 & 0.00266 & 0.0016 & 0.00674 \\
0.0032 & 0.0027 & 0.00093 & 0.00683 \\
0.00152 & 0.00418 & 0.00123 & 0.00693 \\
0.00041 & 0.00326 & 0.00336 & 0.00703 \\
0.00356 & 0.00173 & 0.00183 & 0.00712 \\
0.00426 & 0.00219 & 0.00076 & 0.00721 \\
0.00575 & 0.00018 & 0.00138 & 0.00731 \\
0.00118 & 0.00609 & 0.00014 & 0.00741 \\
0.00305 & 0.00275 & 0.00171 & 0.00751 \\
0.00241 & 0.00302 & 0.00218 & 0.00761 \\
0.0032 & 0.00129 & 0.0032 & 0.00769 \\
0.00055 & 0.00724 & 0.0 & 0.00779 \\
0.001 & 0.00239 & 0.00451 & 0.0079 \\
0.00021 & 0.00335 & 0.00443 & 0.00799 \\
0.00166 & 0.00316 & 0.00326 & 0.00808 \\
0.00533 & 0.00225 & 0.00059 & 0.00817 \\
0.00105 & 0.00577 & 0.00146 & 0.00828 \\
0.0026 & 0.00442 & 0.00135 & 0.00837 \\
0.00032 & 0.00373 & 0.00441 & 0.00846 \\
0.00304 & 0.003 & 0.00252 & 0.00856 \\
0.00099 & 0.00254 & 0.00512 & 0.00865 \\
0.00096 & 0.00164 & 0.00615 & 0.00875 \\
0.00406 & 0.00333 & 0.00146 & 0.00885 \\
0.00224 & 0.00409 & 0.00261 & 0.00894 \\
0.00021 & 0.00311 & 0.00572 & 0.00904 \\
0.00367 & 0.00328 & 0.00218 & 0.00913 \\
0.00518 & 8e-05 & 0.00398 & 0.00924 \\
0.00168 & 0.00467 & 0.00298 & 0.00933 \\
0.00432 & 0.0005 & 0.0046 & 0.00942 \\
0.00338 & 0.00348 & 0.00267 & 0.00953 \\
0.00132 & 0.00707 & 0.00123 & 0.00962 \\
9e-05 & 0.00701 & 0.00261 & 0.00971 \\
0.00301 & 0.00356 & 0.00324 & 0.00981 \\
0.00304 & 0.00147 & 0.00539 & 0.0099 \\
0.00151 & 0.00682 & 0.00167 & 0.01 \\
\hline
\end{tabular}
\end{minipage}
\caption{Pauli X, Y, and Z rates for the Pauli-Lindblad generators used on the Pauli quantum error channel generated by Pauli Lindblad dissipators across the 100 samples reported in Fig.~\ref{fig_allnoisemodels}d.}
\label{tab:pauliLinbland-probabilities}
\end{table}

\section{Additional experimental results with PIE}
\begin{figure*}
\includegraphics[width=1.0\linewidth]{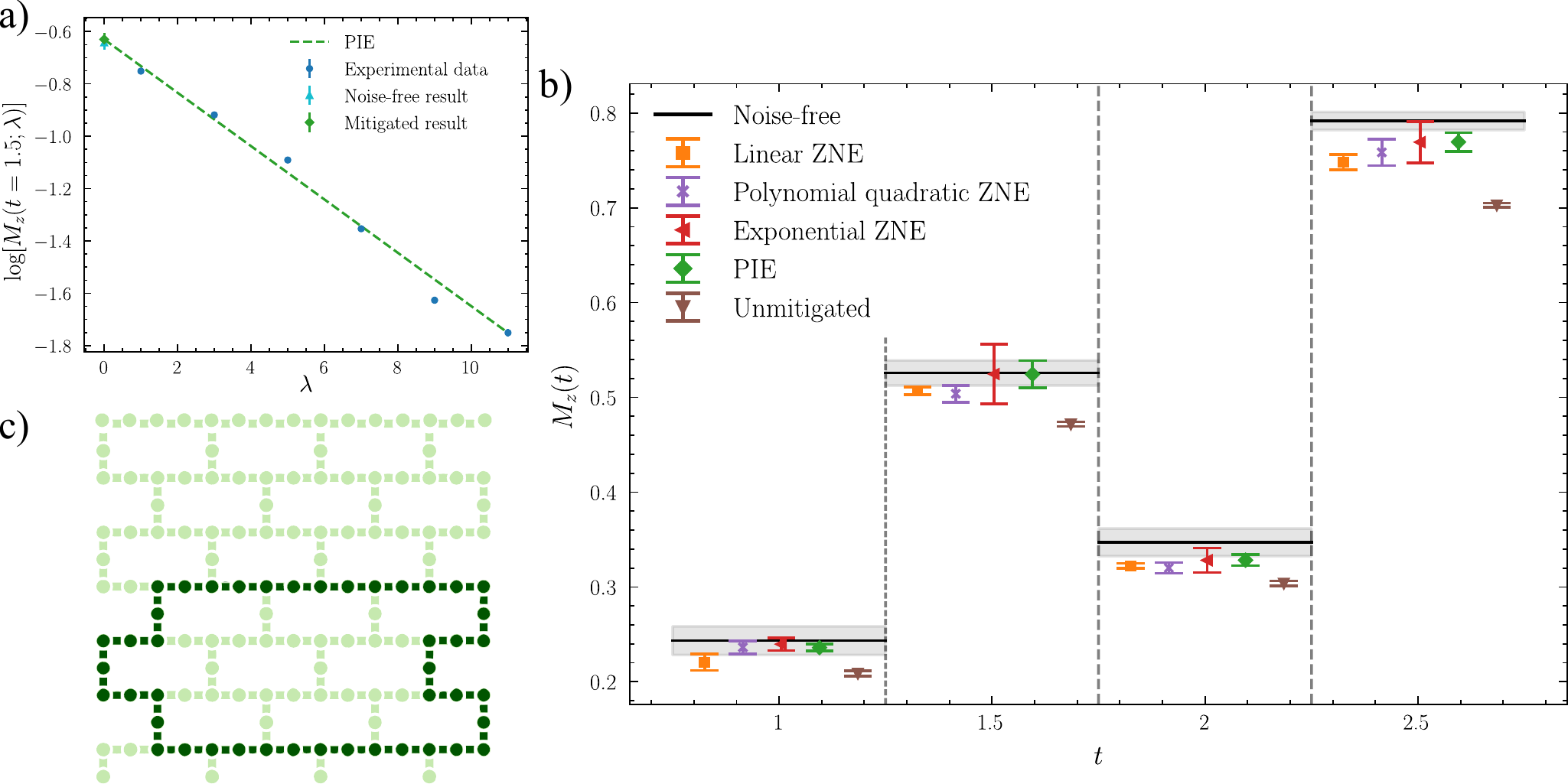}
\caption{\label{fig_experimentalresultsN44} (Color online) Experimental error-mitigated global magnetization $M_z$ on the two Trotter steps simulation of the one-dimensional Ising chain with $N=44$ spins. (a) PIE extrapolation from the noise amplified expectation values for $t=1.5$. (b) Comparison of extrapolation methods. (c) Topology of the IBM Heron superconducting processor ibm torino with the ring of qubits selected for the experiments in dark green. All experiments were performed using six extrapolation points and 4096 shots.}
\end{figure*}

\begin{figure*}
\includegraphics[width=1.0\linewidth]{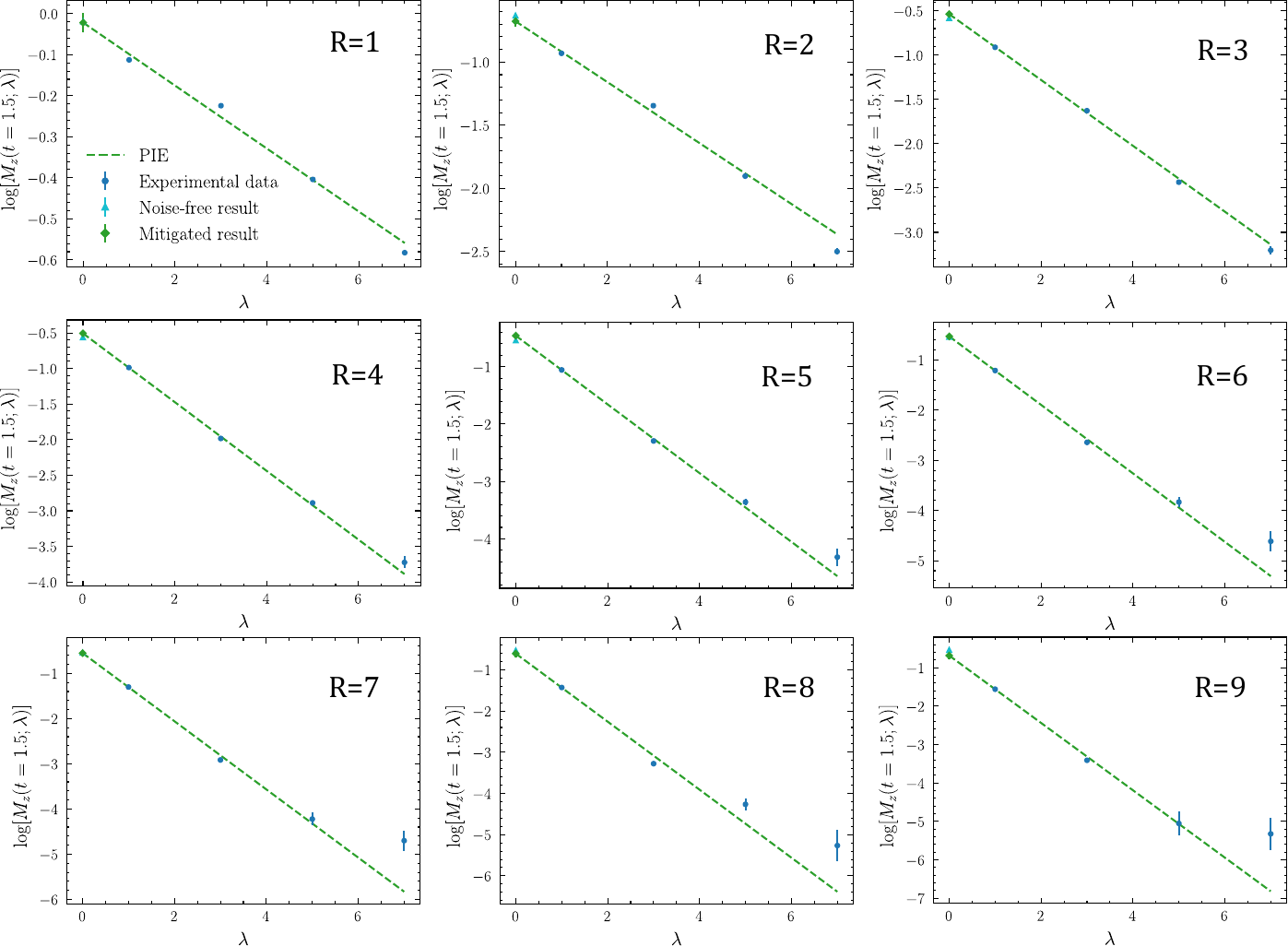}
\caption{\label{fig_experimentalresultsN44_extrapolations_diffdepth} (Color online) PIE extrapolation performed to obtain the error mitigated experimental results of the global magnetization $M_z$ of an Ising $N=84$ chain at time 1.5 shown in Fig.3 of the manuscript. $R$ denotes the number of Trotter steps. The experimental data were obtained from runs on the IBM Eagle ibm kyiv superconducting processor using 4096 shots. To improve the quality of the extrapolation, the experimental points are weighted by the inverse of their standard deviation in the fit.}
\end{figure*}

In this section, we present additional experimental results that have not been reported in the manuscript. 
In~\figref{fig_experimentalresultsN44} we show the experimental error-mitigated global magnetization of the one-dimensional quantum Ising spin chain Hamiltonian with periodic boundary condition and $N=44$ spins at different times $t = 1,\, 1.5,\, 2,$ and $2.5$. 
The time dynamics simulation is approximated by $R=2$ Trotter steps quantum circuits executed on the IBM Heron superconducting processor \textit{ibm torino}. Extrapolations are performed by erroneous expectation values at 6 noise levels ($\lambda=1,3,5,7,9,11$) computed from 4096 shots, with readout error mitigation applied via IBM's \textit{twirled readout error extinction} protocol. We obtain results that are consistent with the conclusions of other experiments described in the manuscript. 
In~\figref{fig_experimentalresultsN44_extrapolations_diffdepth} we show the experimental data and extrapolations performed at different Trotter steps to evaluate PIE performance at increasing quantum circuit depth in Fig.3 of the manuscript. Experimental data were obtained from runs on the IBM Eagle ibm kyiv superconducting processor using 4096 shots and circuit folding to obtain erroneous estimates of the global magnetization $M_z$ of the $N=84$ quantum Ising spin chain at noise levels $\lambda=1,3,5,7$.

\section{Comparison of PIE with learning-based mitigation methods}
\label{sec_comparisonlearningbased}

In recent years, the application of machine learning techniques to quantum error mitigation has been explored in various ways. Learning-based quantum error mitigation methods leverage classical statistical models to infer noise-free expected values from data generated on noisy quantum hardware. These approaches create flexible mappings between noisy and ideal observables directly from training data obtained through a combination of classical simulation and quantum execution. Machine learning-based quantum error mitigation methods have proven to be scalable tools capable of producing accurate results at low cost once the training phase is complete. 

PIE has advantages over other recent learning-based methods in terms of computational resources, as it does not require a pretrained model or a large amount of data. Furthermore, current learning-based methods are dominated by heuristics that lack the theoretical guarantees or physical insights provided by PIE. However, it remains to be seen how PIE compares to these methods in terms of accuracy.

In this appendix, we compare the results of PIE with one of the most discussed learning-based approaches to date, known as Clifford Data Regression (CDR). CDR estimates ideal expectation values by fitting a regression model to data gathered from classically simulable near‑Clifford circuits. The central idea is that the noisy expectation value of an observable for a given circuit is smoothly related to its ideal, noise‑free value. To exploit this, CDR constructs a training set of circuits $\{\ket{\Psi_t}\}$ derived from the target circuit $\{\ket{\Psi_t}\}$ by replacing non‑Clifford operations by Clifford gates while preserving overall structure. For each such training circuit, the ideal expectation value is computed via efficient classical simulation $\langle\mO\rangle^{\rm ideal}_i$, and the corresponding noisy value is measured on the quantum device $\langle\mO\rangle^{\rm err}_i$. A low‑complexity regression model is then fit to the resulting dataset $\{\langle\mO\rangle^{\rm err}_i,\langle\mO\rangle^{\rm ideal}_i\}$. Once trained, this model is applied to the noisy estimate obtained from the original circuit $\langle\mO\rangle^{\rm err}_t$, yielding a corrected, approximately noise‑free expectation value $\langle\mO\rangle^{\rm ideal}_t$. A key hyperparameter in CDR is the fraction of non-Clifford gates that are retained or replaced by Clifford gates in the training circuits, which we denote here by $f$. The higher the fraction of non-Clifford gates that are retained in the training circuits, the more accurate the model will be, but the more expensive it will be to simulate the training circuits and, therefore, the greater the computational overhead of error mitigation.

\begin{figure*}
\includegraphics[width=0.7\linewidth]{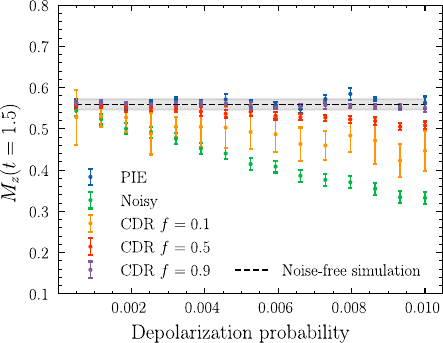}
\caption{\label{fig_PIEcomparisonwithCDR} (Color online) Comparison of PIE and Clifford Data Regression performance. We show the numerical results of the global magnetization $M_z$ in the three-step Trotter simulation of the one-dimensional Ising chain at time $t=1.5$ with $N = 8$ spins using a depolarizing noise model~\eqref{eq_depolnoise}. We show the unmitigated or noisy result (green dot), the noise-free value (dashed line), and the error-mitigated values by PIE (blue dot) and CDR with several fractions of non-Clifford gates in the training circuits $f$ (yellow, red, and purple dots). All results show their corresponding standard deviation as whiskers or a shadow. PIE was performed using four extrapolation points. CDR was implemented using 20 training circuits per expected value and linear extrapolation, and we show the average of 10 executions. All numerical experiments were performed using 4096 shots.}
\end{figure*}

In Fig.\ref{fig_PIEcomparisonwithCDR} we display the numerical results from the computation of the global transverse
field magnetization $M_z$ of the one-dimensional quantum Ising spin chain Hamiltonian~\eqref{eq_globalmagnetization} with $\Gamma=0.5$ and $N=8$ spins at $t = 1.5$. We simulate the time-evolution of the quantum state with 3 Trotter steps under a depolarizing noise channel~\eqref{eq_depolnoise}, and use both PIE and CDR to extract error-mitigated expectation values from noisy results calculated as the average of 4096 shots. We observe that, while the CDR method is capable of matching or even outperforming PIE when $f$ is high (which in practice will imply more computational resources), our results show that PIE offers more robust and accurate predictions than CDR when the fraction of non-Clifford gates is moderate. Therefore, in the low-noise regime relevant to PIE, PIE provides results consistent with the best CDR performance, but avoids the computational overhead produced by the cost of the training process. It should be noted that CDR can be improved by recently proposed methods~\cite{Lowe_2021,Perez_extension_2024}.

\end{document}